%% file: entropy_v2.tex
\RequirePackage{fix-cm}
\RequirePackage{amsmath} 
\pdfoutput=1
\documentclass[smallextended]{svjour3}

\usepackage{setspace}

\smartqed  
\usepackage{graphicx}
\usepackage{tabularray}
\usepackage{mathptmx}      
\usepackage[misc]{ifsym}

\usepackage{xcolor}
\usepackage{url}
\usepackage{hyperref}
\makeatletter
\renewcommand*{\eqref}[1]{%
  \hyperref[{#1}]{\textup{\tagform@{\ref*{#1}}}}%
}
\makeatother

\usepackage[authoryear]{natbib}
\bibpunct{(}{)}{}{a}{}{;}

\usepackage{hyperref}
\usepackage{amsfonts}
\usepackage{amssymb}

\usepackage{enumerate}
\usepackage{bbm} 

\spnewtheorem{mytheorem}{Theorem}{\bf}{\rm}
\spnewtheorem{myproposition}{Proposition}{\bf}{\rm}
\spnewtheorem{mylemma}{Lemma}{\bf}{\rm}
\spnewtheorem{myremark}{Remark}{\bf}{\rm}

\usepackage{apptools}
\AtAppendix{\counterwithin{mytheorem}{section}}
\AtAppendix{\counterwithin{myproposition}{section}}
\AtAppendix{\counterwithin{definition}{section}}
\AtAppendix{\counterwithin{mylemma}{section}}
\AtAppendix{\counterwithin{myremark}{section}}
\AtAppendix{\counterwithin*{equation}{section}}
\AtAppendix{}

\renewcommand{\tens}[1]{\mathrm{#1}}
\renewcommand{\vec}[1]{\mathbf{#1}}

\usepackage{siunitx}
\DeclareSIUnit\year{yr}
\DeclareSIUnit\carbon{C}

\newcommand{\R}{\mathbb{R}}
\newcommand{\N}{\mathbb{N}}
\renewcommand{\P}{\mathbb{P}}
\newcommand{\E}{\mathbb{E}}
\newcommand{\TT}{\mathcal{T}}
\renewcommand{\H}{\mathbb{H}}

\newcommand{\PH}{\operatorname{PH}}
\newcommand{\Exp}{\operatorname{Exp}}
\newcommand{\Poi}{\operatorname{Poi}}

\newcommand{\suml}{\sum\limits}
\newcommand{\prodl}{\prod\limits}
\newcommand{\intl}{\int\limits}
\newcommand{\liml}{\lim\limits}

\newcommand{\deriv}[1]{\frac{\operatorname{d}}{\operatorname{d}#1}}
\newcommand{\dd}[1]{\,\mathrm{d}#1}
\newcommand{\pderiv}[1]{\frac{\partial}{\partial #1}}

\newcommand{\peta}{\mathrm{P}}
\newcommand{\gC}{\mathrm{gC}}
\newcommand{\yr}{\mathrm{yr}}
\newcommand{\meter}{\mathrm{m}}
\newcommand{\bits}{\mathrm{bits}}
\newcommand{\nats}{\mathrm{nats}}

\newcommand{\vnorms}[1]{\|#1\|}
\newcommand{\transpose}{T}
\newcommand{\diag}{\operatorname{diag}}
\newcommand{\ie}{that is}

\newcommand{\pdf}{probability density function}
\newcommand{\NPP}{\ensuremath{\mathrm{NPP}}}

\renewcommand{\emph}[1]{``#1''}

\usepackage{booktabs}
\usepackage{graphicx}
\usepackage{amsmath, amsfonts}
\usepackage{pdflscape}
\usepackage{tikz}

\journalname{Mathematical Geosciences}

\begin{document}

\title{Information content and maximum entropy of compartmental systems in equilibrium
}

\titlerunning{Entropy in compartmental models}        

\author{Holger Metzler \and Carlos A. Sierra}


\institute{Holger Metzler \at
  Max Planck Institute for Biogeochemistry,
  Hans-Kn\"oll-Str. 10,
  07745 Jena,
  Germany.
  Now at:
  Department of Forest Ecology and Management,
  Swedish University of Agricultural Sciences,
  Skogsmarksgr\"and 17,
  901~83 Ume\r{a},
  Sweden.
  In between at:
  Department of Crop Production Ecology,
  Swedish University of Agricultural Sciences,
  Ulls v\"ag 16,
  756~51 Uppsala,
  Sweden,
  \email{holger.metzler@slu.de},
  ORCiD: 0000-0002-8239-1601
  \and
  Carlos A. Sierra \at
  Max Planck Institute for Biogeochemistry,
  Hans-Kn\"oll-Str. 10,
  07745 Jena,
  Germany,
  \email{csierra@bgc-jena.mpg.de}\\
  Department of Ecology,
  Swedish University of Agricultural Sciences,
  Ulls v\"ag 16,
  756~51 Uppsala,
  Sweden,
  ORCiD: 0000-0003-0009-4169
}

\date{Received: \hspace{2cm} Accepted: \hspace{2cm}}

\maketitle

\begin{abstract}
Although compartmental dynamical systems are used in many different areas of science, model selection based on the maximum entropy principle (MaxEnt) is challenging because of the lack of methods for quantifying the entropy for this type of systems. 
Here, we take advantage of the interpretation of compartmental systems as continuous-time Markov chains to obtain entropy measures that quantify model information content. 
In particular, we quantify the uncertainty of a single particle's path as it travels through the system as described by path entropy and entropy rates.
Path entropy measures the uncertainty of the entire path of a traveling particle from its entry into the system until its exit, whereas entropy rates measure the average uncertainty of the instantaneous future of a particle while it is in the system.
We derive explicit formulas for these two types of entropy for compartmental systems in equilibrium based on Shannon information entropy and show how they can be used to solve equifinality problems in the process of model selection by means of MaxEnt.


\section*{Statements and Declarations}
\textbf{Competing Interests:} The authors have no relevant financial or non-financial interests to disclose.\\
\textbf{Code Repository:} The Python code to reproduce the figures used in the manuscript is currently provided at\\
\url{https://github.com/goujou/entropy_and_complexity_in_eq}.\\
In case of publication it will be transformed into a permanent repository with a DOI attached to it.
\end{abstract}

\section{Introduction}\label{intro}
For many modeling applications, it is of interest to quantify the complexity of the system of differential equations used to represent natural phenomena \citep{Burnham2002, Hoege2018}. In principle, we are interested in selecting models that are parsimonious; i.e. with the least degree of complexity for explaining certain patterns in nature \citep{Golan2022}. 
The concept of entropy has been commonly used to characterize complexity or information content. Classical entropy measures for dynamical systems characterize the rate of increase in dynamical complexity as the system evolves over time \citep{Jost2005}. These metrics have been used extensively to characterize chaotic behavior in complex nonlinear systems \citep{Fan2021}, but as we will see later, they give trivial results for a large range of models used in geosciences and biology. 

In a large variety of scientific fields models are based on the principle of mass conservation.
In many cases such models are nonnegative dynamical systems that can be described by first-order systems of ordinary differential equations (ODEs) with strong structural constraints.
Such systems are called compartmental systems \citealp{Anderson1983, Walter1999, Haddad2010}.

Compartmental systems can be evaluated using diagnostic metrics that predict system-level behavior and allow comparisons of systems of very different structures. 
Age and transit time of material content in compartmental systems  are two diagnostic metrics that have been widely studied for systems in and out of equilibrium \citep{Eriksson1971ARoEaS, Bolin1973tellus, Rasmussen2016JMB, Sierra2017GCB, Metzler2018MGS, MetzlerMuellerSierra2018PNAS}.
They help compare behavior and quality of different models.
Nevertheless, structurally very different models might show very similar ages and transit times and might represent equally well a given measurement.
If we are in the position to choose among such models, which is the one to select?
This equifinality problem can be resolved by the maximum entropy principle (MaxEnt) \citep{Jaynes1957PR1, Jaynes1957PR2}, a generic procedure to draw unbiased inferences from measurement or stochastic data \citep{Presse2013RMP, Golan2022}.

In order to apply MaxEnt to compartmental systems, some appropriate notion of entropy is required to measure the system's uncertainty or information content.
Two classical examples in dynamical systems theory are the topological entropy and the Kolmogorov-Sinai/metric entropy.
However, open compartmental systems are dissipative and by Pesin's theorem \citep{Pesin1977UMN} both metric and topological entropy vanish and cannot serve as a measure of uncertainty.
Alternatively, we can interpret compartmental systems as weighted directed graphs.
\citet{Dehmer2011IS} provide a comprehensive overview of the history of graph entropy measures.
Unfortunately, most of such entropy measures are based on the number of vertices, vertex degree, edges, or degree sequence \citep{Trucco1956BoMB, Morzy2017}.
Thus, they concentrate only on the structural information of the graph.
There are also graph theoretical measures that take edges and weights into account by using probability schemes.
Their drawback is that the underlying meaning of uncertainty becomes difficult to interpret because the assigned probabilities seem somewhat arbitrary \citep{Bonchev2005}.

To bridge this gap we interpret deterministic compartmental systems from a probabilistic viewpoint.
Based on the Shannon information entropy \citep{Shannon1949TUoIP} of the continuous-time Markov chain that describes the random path of a single particle through the compartmental system \citep{Metzler2018MGS}, we introduce three entropy measures.
While the path entropy describes the uncertainty of a single particle's path through the system, the entropy rate per unit time and the entropy rate per jump describe average uncertainties over the course of a particle's journey. 
The focus on a single particle makes our entropies microscopic system properties and consequently distinguishes our approach from the theory of maximum caliber \citep[MaxCal,][]{jaynes1985macroscopic, Roach2020}, where path entropy is interpreted as a macroscopic system property of bulk material.
Furthermore, our information theoretical approach differs from the thermodynamic approach to entropy, which has been developed by other authors studying energy transfers and reversibility in thermodynamic systems \citep{Aoki1988, Haddad2010, Haddad2013, Haddad2019}.

The manuscript is organized as follows.
First we provide the fundamentals from information theory and dynamical systems theory that are necessary to introduce the path entropy as the uncertainty of a single particle traveling through the system.
Then, we mathematically derive the path entropy and introduce the entropy rates per unit time and per jump, before we introduce MaxEnt and structural model identification.
Afterwards we present the introduced theory by means of simple generic examples from the field of carbon-cycle modeling exploring the effect of different parameterizations on the three entropy metrics, before we apply MaxEnt to a model identification problem.
Then we discuss the results and draw final conclusions.

\section{Mathematical background: information entropy and compartmental systems as Markov chains}
First, we introduce some basic notations and well-known properties of Shannon information entropy of random variables and stochastic processes.
Then, we present compartmental systems as a means to model material-cycle systems that obey the law of mass balance.
We then consider such systems from a single-particle point of view and define the path of a single particle through the system along with its visited compartments, sojourn times, occupation times, and transit time.

\subsection{Short summary of Shannon information entropy}
\label{sec:entropy_basics}

We introduce a few basic concepts of information entropy.
Within the framework of this manuscript, discrete entropies are usually associated to a particle's jump into another compartment and differential entropies to a particle's sojourn time within a specific compartment.
Entropy rates are defined as average uncertainties of the particle's path while it is in the system.
See Sects.~2 and~8 of \citet{Cover2006} for a more detailed introduction to Shannon's information entropy and differential entropy.
Entropy rates for discrete- and continuous-time stochastic processes are introduced in \citet[Sect.~4][]{Cover2006} and \citet{Dumitrescu1988MICAS}. 

Let $Y$ be a real-valued discrete (continuous) random variable and call $p$ its probability mass function (\pdf).
Then
\begin{equation}
  \label{eqn:entropy}
  \H(Y) := -\E\left[\log p(Y)\right]
\end{equation}
is called the \emph{Shannon information entropy} (\emph{differential entropy}) of $Y$.
Most of the time we just say \emph{entropy} and the precise meaning can be derived from the context.
The entropy's unit depends on the logarithmic base.
For base $2$ the unit is \emph{$\bits$} and for the natural logarithm with base $e$ the unit is \emph{$\nats$}.
Throughout this manuscript we use the latter if not stated otherwise.

The entropy $\H(Y)$ of a random variable $Y$ has two intertwined interpretations.
On the one hand, it is a measure of uncertainty, \ie, a measure of how difficult it is to predict the outcome of a realization of $Y$.
On the other hand, $\H(Y)$ is also a measure of the information content of $Y$, \ie, a measure of how much information we gain once we learn about the outcome of a realization of $Y$.
It is important to note that, even though their definitions and information theoretical interpretations are quite similar, the Shannon- and the differential entropy have one main difference.
The Shannon entropy is always nonnegative, whereas the differential entropy can have negative values.
While the Shannon entropy is an absolute measure of information and makes sense in its own right, the differential entropy is not an absolute information measure, is not scale-invariant, and makes sense only in comparison with the differential entropy of another random variable.

\begin{figure}[htbp]
  \vspace{-0.6cm}
  \centering
  \includegraphics[width=1.0\linewidth]{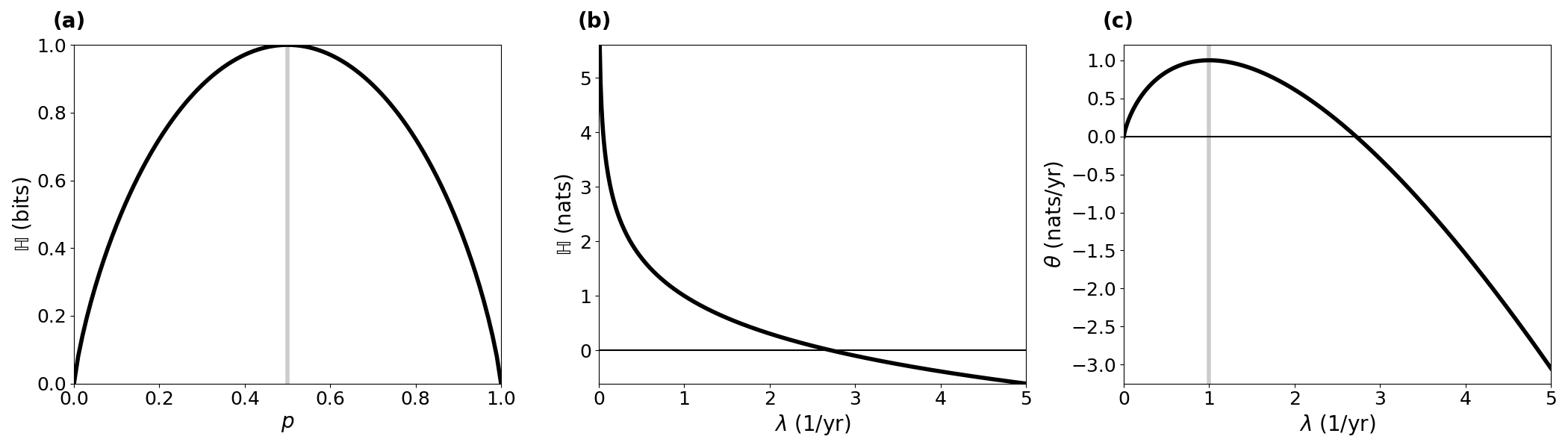}
  \caption{
  Shannon entropy of a Bernoulli distribution (a), differential entropy of an exponential distribution (b), and entropy rate of a Poisson process (c).
  Vertical gray lines indicate the parameter values leading to the highest entropy
  }
  \label{fig:simple_entropy}
\end{figure}

Panel (a) of Fig.~\ref{fig:simple_entropy} depicts the Shannon entropy with logarithmic base $2$ of a Bernoulli random variable $Y$ with $\P(Y=1)=1-\P(Y=0)=p\in[0,1]$ representing a coin toss with probability of heads equal to $p$.
The closer $p$ is to $1/2$ the more difficult it is to predict the outcome, and for an unbiased coin with $p=1/2$ we have no information about the outcome whatsoever, and the Shannon entropy
\begin{equation}
	\H(Y) = -p\,\log p - (1-p)\,\log(1-p)
\end{equation}
is maximized.
Panel (b) of Fig.~\ref{fig:simple_entropy} shows the differential entropy of an exponentially distributed random variable $Y\sim\Exp(\lambda)$ with rate parameter $\lambda>0$, \pdf\ $f(y) = \lambda\,e^{-\lambda\,y}$ for $y\geq0$, and $\E\left[Y\right]=\lambda^{-1}$.
We can imagine it to represent the duration of stay of a particle in a well-mixed compartment in an equilibrium compartmental system, where $\lambda$ is the total outflow rate from the compartment.
The higher the outflow rate is, the more likely an early exit of the particle is, and the easier it is to predict its moment of exit.
Hence, the differential entropy 
\begin{equation}
	\H(Y) = 1-\log\lambda
\end{equation}
decreases with increasing $\lambda$.

The \emph{joint entropy} of two random variables $Y_1$ and $Y_2$ can be described as
\begin{equation}
  \H(Y_1,Y_2) = \H(Y_1) + \H(Y_2\,|\,Y_1),
\end{equation}
where the \emph{conditional entropy} $\H(Y_2\,|\,Y_1)$ describes the uncertainty of $Y_2$ under the condition that $Y_1$ is known.
The uncertainty of a stochastic process $Z$ can be measured by its \emph{entropy rate} $\theta(Z)$, which describes the time density of the average information in the process.
For instance, let $Z\sim\Poi(\lambda)$ be a Poisson process with intensity rate $\lambda>0$ describing the moments of occurrence of certain events.
The interarrival times of $Z$ or the times between events are $\Exp(\lambda)$-distributed, such that in the long run on average the time span between events has length $\lambda^{-1}$.
The entropy of the interarrival times is given by $\H(\Exp(\lambda))=1-\log \lambda$, and averaging it over the mean interarrival time gives the entropy rate of the Poisson process $Z$ \citep[Sect.~3.3]{Gaspard1993PR}, \ie,
\begin{equation}
  \theta(Z) = \theta(\Poi(\lambda)) = \lambda\,(1-\log \lambda).
\end{equation}
This entropy rate increases with $\lambda\in[0,\,1]$, reaches its maximum at $1$, and then it decreases (Fig.~\ref{fig:simple_entropy}, panel c).
This behavior is independent of the unit of $\lambda$, because it is based on the differential entropy of the exponential distribution and hence not scale-invariant.
Consequently, it is not an absolute measure of information content, but only useful in comparison to the entropy rates of other stochastic processes.

\subsection{Compartmental systems in equilibrium}\label{sec:one_particle}
Mass-balanced flow of material into a system, within the system and out of the system that consists of several compartments can be modeled by so-called compartmental systems \citep{Anderson1983, Jacquez1993SIAM}.
Compartments are always well-mixed and usually also called \emph{pools} or \emph{boxes}.
An autonomous compartmental systems can be described by the $d$-dimensional linear ODE system
\begin{equation}\label{eqn:lin_CS_sys}
  \deriv{t}\,\vec{x}(t) = \tens{B}\,\vec{x}(t) + \vec{u},\quad t>0,\\
\end{equation}
with some nonnegative initial condition $\vec{x}(0) = \vec{x}^0\in\R^d_+$.
The nonnegative vector $\vec{x}(t)$ describes the amount of material in the different compartments at time $t$, the nonnegative vector $\vec{u}=(u_i)_{i=1,2,\ldots,d} \in\R^d_+$ is the vector of external inputs to the compartments, and the compartmental matrix $\tens{B}\in\R^{d\times d}$ describes the flux rates between the compartments and out of the system.
The nonnegative off-diagonal value $B_{ij}$ is the flux rate from compartment $j$ to compartment $i$, the absolute value of the negative diagonal value $B_{jj}$ is the total rate of fluxes out of compartment $j$, and the nonnegative value $z_j=-\sum_{i=1}^d B_{ij}$ is the rate of the flux from compartment $j$ out of the system.
By requiring $\tens{B}$ to be invertible we ensure that the system is \emph{open}, \ie, all material that enters the system will eventually also leave it.
Throughout this manuscript, we consider the open compartmental system \eqref{eqn:lin_CS_sys} to have reached its unique steady-state or equilibrium compartment vector $\vec{x}^\ast = -\tens{B}^{-1}\,\vec{u}$.
This implies $\vnorms{\vec{r}} = \vnorms{\vec{u}}$, where $\vec{r} = (r_i)_{i=1,2,\ldots,d}$ given by $r_j=z_j\,x^\ast_j$ is the external outflux vector from the system, and $\vnorms{\cdot}$ denotes the sum of absolute values of a vector ($l_1$-norm). 
An open compartmental system in equilibrium given by Eq.~\eqref{eqn:lin_CS_sys} is fully characterized by $\vec{u}$ and $\tens{B}$, and we denote it by $M:=M(\vec{u},\tens{B})$.

\subsection{The one-particle perspective}
\label{sec:the_one_particle_perspective}
While Eq.~\eqref{eqn:lin_CS_sys} describes the movement of bulk material through the system, compartmental systems in equilibrium can also be described probabilistically by considering the random path of a single particle through the system \citep{Metzler2018MGS}.
If $X_t\in S:=\{1,2,\ldots,d\}$ denotes the compartment in which the single particle is at time $t$, and $X_t=d+1$ if the particle has already left the system, then $X:=(X_t)_{t\geq0}$ is an absorbing continuous-time Markov chain \citep{Norris1997} on $\widetilde{S}:=S\cup\{d+1\}$.
Its initial distribution is given by $\widetilde{\vec{\beta}}=(\beta_1, \beta_2, \ldots, \beta_d, 0)^T$, where $\vec{\beta}:=\vec{u}/\vnorms{\vec{u}}$, and hence $\beta_j=\P(X_0=j)$ is the probability of the single particle to enter the system through compartment $j$.
The superscript $T$ denotes the transpose of the respective vector or matrix. 
The transition-rate matrix of $X$ is given by
\begin{equation}\label{eqn:Q}
  \tens{Q} =
  \begin{pmatrix}
    \tens{B} & \vec{0} \\
    \vec{z}^T & 0
  \end{pmatrix},
\end{equation}
and thus
\begin{equation}
  \P(X_t=i) = (e^{t\,\tens{Q}}\,\widetilde{\vec{\beta}})_i = \suml_{j=1}^d (e^{t\,\tens{Q}})_{ij}\,\beta_j, \quad i\in\widetilde{S},
\end{equation}
is the probability of the particle to be in compartment $i$ at time $t$ if $i\in S$ or that the particle has left the system if $i=d+1$.
Here, $e^{t\,\tens{Q}}$ denotes the matrix exponential.
Furthermore, 
\begin{equation}
  \P(X_t=i\,|\,X_s=j) = (e^{(t-s)\,\tens{Q}})_{ij},\quad s\leq t,\quad i,j\in\widetilde{S},
\end{equation}
is the probability that $X$ is in state $i$ at time $t$ given it was in state $j$ at time $s$.
Since the Markov chain $X$ and the compartmental system in equilibrium given by Eq.~\eqref{eqn:lin_CS_sys} are equivalent, we can write
\begin{equation}
  M=M(\vec{u},\tens{B}) = M(X).
\end{equation}

\subsection{The path of a single particle}
A particle's path through the system from the moment of entering until the moment of exit can be described as a sequence of (compartment, sojourn-time)-pairs
\begin{equation}
  \label{eqn:path}
  \mathcal{P}(X) := ((Y_1=X_0, T_1),(Y_2,T_2),\ldots,(Y_{\mathcal{N}-1},T_{\mathcal{N}-1}), Y_{\mathcal{N}}=d+1),
\end{equation}
where $X$ is the absorbing Markov chain associated to the particle's journey.
The sequence $Y_1,Y_2,\ldots,Y_{\mathcal{N}-1}\in S$ represents the successively visited compartments  with the associated sojourn times $T_1,T_2,\ldots,T_{\mathcal{N}-1}$, the random variable
\begin{equation}
  \mathcal{N}:=\inf\,\{n\in\N:\,Y_n=d+1\}
\end{equation}
denotes the first hitting time of the absorbing state $d+1$ by the \emph{embedded jump chain} $Y:=(Y_n)_{n=1,2,\ldots,\mathcal{N}}$ of $X$ \citep{Norris1997}.
With $\lambda_j:=-Q_{jj}$ the one-step transition probabilities of $Y$ are given by, for $i,j\in\widetilde{S}$,
\begin{equation}\label{eqn:P_ij}
  P_{ij}:=\P(Y_{n+1}=i\,|\,Y_n=j) = 
  \begin{cases}
    0,\quad & i=j\text{ or }\lambda_j=0,\\
    Q_{ij}/\lambda_j,\quad & \text{else}.
  \end{cases}
\end{equation}
Let $\tens{P}|_S=(P_{ij})_{i,j\in S}$ be the restriction of $\tens{P}$ to $S$.
We can also write $\tens{P}|_S = \tens{B}\,\tens{D}^{-1} + \tens{I}$, where $\tens{D}:= \diag\,(\lambda_1,\lambda_2,\ldots,\lambda_d)$ is the diagonal matrix with the diagonal entries of $\tens{B}$, and $\tens{I}$ denotes the identity matrix of appropriate dimension.
Then $\tens{M}:=(\tens{I}-\tens{P}|_S)^{-1}$ is the \emph{fundamental matrix} of $Y$.
The entry $M_{ij}$ denotes the expected numbers of visits to compartment $i$ given that the particle entered the system through compartment $j$.
Consequently, the expected number of visits to compartment $i\in S$ is given by 
\begin{equation}
  \label{eqn:N_i}
  \E\left[N_i\right] = \suml_{j=1}^d M_{ij}\,\beta_j = (\tens{M}\,\vec{\beta})_i = \left[(\tens{I}-\tens{P}|_S)^{-1}\,\vec{\beta}\right]_i 
  = (\tens{D}\,\tens{B}^{-1}\,\vec{\beta})_i
  = \frac{\lambda_i\,x^\ast_i}{\vnorms{\vec{u}}}
\end{equation}
and the total expected number of jumps is given by
\begin{equation}
  \E\left[\mathcal{N}\right] = \suml_{i=1}^d (\tens{M}\,\vec{\beta})_i+1= \suml_{i=1}^d \E\left[N_i\right]+1,
\end{equation}
where we take into account also the last jump out of the system.

The last jump, $\mathcal{N}$, leads the particle out of the system such that at the moment of this last jump $X$ takes on the value $d+1$.
This last jump happens at the absorption time of the Markov chain $X$, which is defined as
\begin{equation}
   \TT := \inf\,\{t>0: X_t=d+1\}.
\end{equation}
The absorption time is phase-type distributed \citep{Neuts1981}, $\TT\sim\PH(\vec{\beta},\tens{B})$, with \pdf
\begin{equation}
  f_{\TT}(t) = \vec{z}^T\,e^{t\,\tens{B}}\,\vec{\beta},\quad t\geq0.
\end{equation}
It can be shown \citep[Sect.~3.2]{Metzler2018MGS} that the mean or expected value of $\TT$ equals the turnover time \citep{Sierra2017GCB} of system \eqref{eqn:lin_CS_sys} in equilibrium and is given by total stocks over total fluxes, \ie, 
\begin{equation}
  \E\left[\TT\right] = \frac{\vnorms{\vec{x}^\ast}}{\vnorms{\vec{u}}}.
\end{equation}
Furthermore, by construction $\sum_{k=1}^{\mathcal{N}-1} T_k = \TT$.
If we denote by $\mathbbm{1}_{\{A\}}$ the indicator function of the logical expression $A$, given by
\begin{equation}
  \mathbbm{1}_{\{A\}} =
  \begin{cases}
    1, \quad A\text{ is true},\\
    0, \quad\text{else},
  \end{cases}
\end{equation}
then $O_j:=\sum_{k=1}^{\mathcal{N}-1} \mathbbm{1}_{\{Y_k=j\}}\,T_k$ is the total time that the particle spends in compartment $j$.
This time is called \emph{occupation time} of $j$ and its mean is given by \citep[Sect.~3.3]{Metzler2018MGS}
\begin{equation}
  \label{eqn:occupation_time}
  \E\left[O_j\right] = \frac{x^\ast_j}{\vnorms{\vec{u}}},
\end{equation}
which induces $\E\left[\TT\right] = \sum_{j=1}^d \E\left[O_j\right]$.

\section{Entropy measures, MaxEnt, and structural model identification}
Based on these basic structures of the path of a single particle traveling through the system, we compute three different types of entropy, for which we provide below
a summary of the desirable relations among them:
\begin{enumerate}[(1)]
  \item 	As a particle travels through a system of interconnected compartments, it jumps a certain number of times to the next compartment until it finally jumps out of the system.
	Between two jumps, the particle resides in some compartment.
  The \emph{path entropy} measures the entire uncertainty about the particles travel through the system, including both the sequence of visited compartments and the respective times spent there.

	\item The entire travel of the particle takes a certain time.
	In each unit time interval before the particle leaves the system, uncertainties exist whether the particle jumps, where it jumps, and even how often it jumps.
	The mean of these uncertainties over the mean length of the travel interval is measured by the \emph{entropy rate per unit time}.

	\item Each jump comes with uncertainties about which compartment will be next and how long will the particle stay there.
	The \emph{entropy rate per jump} measures the average of these uncertainties with respect to the mean number of jumps until the particle's exit from the system.
\end{enumerate}

Once these entropy metrics are established, we introduce MaxEnt and show how to apply it to the problem of structural model identification.

\subsection{Path entropy, entropy rate per unit time, and entropy rate per jump}
\label{sec:path_entropy}
The path $\mathcal{P}=\mathcal{P}(X)$ given by Eq.~\eqref{eqn:path} can be interpreted in three different ways.
Each of these ways leads to a different interpretation of the path's entropy.
First, we can look at $\mathcal{P}$ as the result of bookkeeping of the absorbing continuous-time Markov chain $X$, where for all times $t$ we note down the pair $(X_t,t)$ of the current compartment and the current time.
Second, we can consider the path as a discrete-time process.
In each time step $n$, we choose randomly a new compartment $Y_{n+1}$ and an associated sojourn time $T_{n+1}$ of the particle in this compartment.
Third, we can look at $\mathcal{P}$ as a single random variable with values in the space of all possible paths.
Based on the latter interpretation we now derive the path entropy.

We are interested in the uncertainty/information content of the path $\mathcal{P}(X)$ of a single particle.
Along the lines of \citet{Albert1962AMS}, we construct a space $\wp$ that contains all possible paths that can be taken by a particle that runs through the system until it leaves.
Let $\wp_n:=(S\times\R_+)^n\times\{d+1\}$ denote the space of paths that visit $n$ compartments/states before ending up in the environmental compartment/absorbing state $d+1$.
By $\wp:=\bigcup_{n=1}^{\infty}\wp_n$ denote the space of all eventually absorbed paths.
Note that, since $\tens{B}$ is invertible, a path through the system is finite with probability $1$.
Let $l$ denote the Lebesgue measure on $\R_+$ and $c$ the counting measure on $S$.
Furthermore, let $\sigma_n$ be the $\sigma$-finite product measure on $\wp_n$.
It is defined by $\sigma_n:=(c\otimes l)^n \otimes c$.
Almost all sample functions of $(X_t)_{t\geq0}$ can be represented as a point $p\in\wp$ \citep[Chapter~VI]{Doob1953}.
Consequently, we can represent $X$ by a finite-length path $\mathcal{P}(X)=((Y_1,T_1),(Y_2,T_2),\ldots,(Y_n,T_n),Y_{n+1})$ for some $n\in\N$, where $Y_{n+1}=d+1$.

For each set $W\subseteq\wp$ for which $W\cap \wp_n$ is $\sigma_n$-measurable for each $n\in\N$, we define $\sigma^\ast(W) := \sum_{n=1}^{\infty} \sigma_n(W\cap\wp_n)$.
This measure is defined on the $\sigma$-field $\mathcal{F}^\ast$ which is the smallest $\sigma$-field containing all sets $W\subseteq\wp$ whose projection on $\R^n_+$ is a Borel set for each $n\in\N$.
Let $\sigma$ be a measure on all sample functions, defined for all subsets $W$ whose intersection with $\wp$ is in $\mathcal{F}^\ast$. 
We define it by $\sigma(W):=\sigma^*(W\cap\wp)$.

Let $p=((x_1,t_1),(x_2,t_2,),\ldots,(x_n,t_n),d+1)\in\wp$ for some $n\in\N$.
For $i\neq j$, denote by $N_{ij}(p)$ the total number of path $p$'s one-step transitions from $j$ to $i$ and by $R_j(p)$ the total amount of time spent in $j$.

\begin{mytheorem}\label{theorem:path_pdf}
	The \pdf\ of $\mathcal{P}=\mathcal{P}(X)$ with respect to $\sigma$ is given by
	\begin{equation}
    \begin{aligned}
      f_{\mathcal{P}}(p) = \beta_{x_1}\Bigg(\prodl_{j=1}^d\,\prodl_{i=1,i\neq j}^{d+1} &(Q_{ij})^{N_{ij}(p)}\Bigg)\prodl_{j=1}^d e^{-\lambda_j\,R_j(p)},\\
      & p=((x_1,t_1),(x_2,t_2),\ldots,(x_n,t_n),d+1)\in\wp.
    \end{aligned}
	\end{equation}
\end{mytheorem}

\begin{proof}
	Let $x_1,x_2,\ldots,x_n\in S$, $x_{n+1}=d+1$, and $t_1,t_2,\ldots,t_n\in\R_+$.
	Since
	\begin{equation}
    \begin{aligned}
      &\P((Y_1=x_1,T_1\leq t_1),\ldots,\,(Y_n=x_n,T_n\leq t_n),\, Y_{n+1}=d+1) \\
      &\qquad= \P(Y_{n+1}=d+1\,|\,Y_n=x_n)\\
      &\qquad\qquad\cdot\,\prodl_{k=2}^n \P(Y_k=x_k,T_k\leq t_k\,|\,Y_{k-1}=x_{k-1})\,\P(Y_1=x_k,T_1\leq t_1)\\
      &\qquad= P_{d+1,x_n}\left[\prodl_{k=2}^n P_{x_{k} x_{k-1}}\left(1-e^{-\lambda_{x_k}\,t_k}\right)\right] \beta_{x_1}\left(1-e^{-\lambda_{x_1}\,t_1}\right)\\
      &\qquad= \intl_{\mathbb{T}_n} \beta_{x_1}\prodl_{k=1}^n Q_{x_{k+1}x_k}\,e^{-\lambda_{x_k}\,\tau_k}\,\mathrm{d}\tau_1\mathrm{d}\tau_2\cdots\mathrm{d}\tau_n
    \end{aligned}
  \end{equation}
	with $\mathbb{T}_n=\{(\tau_1,\tau_2,\ldots,\tau_n)\in\R^n_+:\,0\leq\tau_1\leq t_1,0\leq\tau_2\leq t_2,\ldots,0\leq\tau_n\leq t_n\}$,
	the \pdf\ of $\mathcal{P}=\mathcal{P}(x)$ with respect to $\sigma$ is given by
	\begin{equation}
    \begin{aligned}
      f_{\mathcal{P}}(p) = \beta_{x_1}\prodl_{k=1}^n &Q_{x_{k+1}x_k}\,e^{-\lambda_{x_k}\,t_k},\\
      & p=((x_1,t_1),(x_2,t_2),\ldots,(x_n,t_n),d+1)\in\wp.
    \end{aligned}
  \end{equation}
	The term $Q_{x_{k+1}x_k}=Q_{ij}$ enters exactly $N_{ij}(p)$ times.
	Furthermore,
	\begin{equation}
    \begin{aligned}
      \prodl_{k=1}^n e^{-\lambda_{x_k}\,t_k} &= \prodl_{k=1}^n\,\prodl_{j=1}^d \mathbbm{1}_{\{x_k=j\}}\,e^{-\lambda_j\,t_k}
      = \prodl_{j=1}^d e^{-\lambda_j\,\suml_{k=1}^n \mathbbm{1}_{\{x_k=j\}}\,t_k}\\
      &= \prodl_{j=1}^d e^{-\lambda_j\,R_j(p)}.
    \end{aligned}
  \end{equation}
	We make the according substitutions and the proof is finished.
\end{proof}

The entropy of the absorbing continuous-time Markov chain $X$ is equal to its entropy on the random but finite time horizon $[0,\,\TT]$, which in turn equals the entropy of a single particle's path $\mathcal{P}$ through the system.

\begin{mytheorem}\label{thm:entropy_of_X}
	The entropy of the absorbing continuous-time Markov chain $X$ is given by
	\begin{equation}
    \label{eqn:H_X}
    \begin{aligned}
      \H(X) &= \H(\mathcal{P})\\
      &= -\suml_{i=1}^d\beta_i\,\log\beta_i\\
      &\quad + \suml_{j=1}^d \frac{x^\ast_j}{\vnorms{\vec{u}}}\left[\suml_{i=1,i\neq j}^d \,B_{ij}\,(1-\log B_{ij}) + z_j\,(1-\log z_j)\right].
    \end{aligned}
	\end{equation}
\end{mytheorem}

\begin{proof}
	Let $X$ have the finite path representation 
	\begin{equation}
		\mathcal{P}=\mathcal{P}(X)=((Y_1,T_1),(Y_2,T_2),\ldots,(Y_n,T_n),d+1)
	\end{equation}
	for some $n\in\N$, and denote by $f_{\mathcal{P}}$ its \pdf.
	Then, by Theorem~\ref{theorem:path_pdf},
	\begin{equation}
		-\log f_{\mathcal{P}}(\mathcal{P}) = -\log\beta_{Y_1} - \suml_{j=1}^d\,\suml_{i=1,i\neq j}^{d+1}N_{ij}(\mathcal{P})\,\log Q_{ij} + \suml_{j=1}^d \lambda_j\,R_j(\mathcal{P}).
	\end{equation}
	We compute the expectation and get
	\begin{equation}
    \begin{aligned}
      \H(X) &= \H(\mathcal{P}) = -\E\left[\log f_{\mathcal{P}}(\mathcal{P})\right]\\
      &= -\E\left[\log\beta_{Y_1}\right] - \suml_{j=1}^d\,\suml_{i=1,i\neq j}^{d+1}\E\left[N_{ij}(\mathcal{P})\right]\,\log Q_{ij} + \suml_{j=1}^d \lambda_j\,\E\left[R_j(\mathcal{P})\right]\\
      &= \H(Y_1) + \suml_{j=1}^d \lambda_j\,\E\left[R_j(\mathcal{P})\right] - \suml_{j=1}^d\,\suml_{i=1,i\neq j}^{d+1}\E\left[N_{ij}(\mathcal{P})\right]\,\log Q_{ij}.
    \end{aligned}
  \end{equation}
	Obviously, $\E\left[R_j(\mathcal{P})\right]=\E\left[O_j\right]=x^\ast_j/\vnorms{\vec{u}}$ is the mean occupation time of compartment $j\in S$ by $X$.
	Furthermore, for $i\in\widetilde{S}$ and $j\in S$ such that $i\neq j$, by Eqs.~\eqref{eqn:N_i} and~\eqref{eqn:P_ij},
	\begin{equation}
		\E\left[N_{ij}(\mathcal{P})\right] = \E\left[N_j(\mathcal{P})\right]\,P_{ij} = 
		\begin{cases}
			\frac{x^\ast_j}{\vnorms{\vec{u}}}\,B_{ij},\quad & i\leq d,\\
			\frac{x^\ast_j}{\vnorms{\vec{u}}}\,z_j,&i=d+1.
		\end{cases}
	\end{equation}
	Together with $\lambda_j=\sum_{i=1,i\neq j}^{d} B_{ij}+z_j$, we obtain
	\begin{equation}
    \begin{aligned}
      \H(X) &= \H(Y_1)+ \suml_{j=1}^d \frac{x^\ast_j}{\vnorms{\vec{u}}}\Bigg[\left(\suml_{i=1,i\neq j}^d B_{ij}+z_j\right)\\
      &\qquad\qquad\qquad\qquad - \suml_{i=1,i\neq j}^d B_{ij}\,\log B_{ij} - z_j\,\log z_j\Bigg]\\
      &= -\suml_{i=1}^d \beta_i\log\beta_i + \suml_{j=1}^d \frac{x^\ast_j}{\vnorms{\vec{u}}} \Bigg[\suml_{i=1,i\neq j}^d B_{ij}\,(1-\log B_{ij})\\
      &\qquad\qquad\qquad\qquad\qquad + z_j\,(1-\log z_j)\Bigg].
    \end{aligned}
  \end{equation}
\end{proof}

By some simple substitutions and rearrangements, we obtain two representations of $\H(X)=\H(\mathcal{P})$ that are easy to interpret.
For simplicity of notation, we define
\begin{equation}
  \H(\vec{\beta}) := -\suml_{i=1}^d\beta_i\,\log\beta_i.
\end{equation}

\begin{myproposition}\label{prop:entropy_of_X}
	The entropy of the absorbing continuous-time Markov chain $X$ is also given by
	\begin{equation}
	  \label{eqn:H_occupation_time}
	  \H(X) = \H(\vec{\beta}) + \suml_{j=1}^d \E\left[O_j\right] \left(\suml_{i=1,\,i\neq j}^d \theta(\Poi(B_{ij})) + \theta(\Poi(z_j))\right)
	\end{equation}
	and
	\begin{equation}
	  \label{eqn:H_number_of_visits}
	  \begin{aligned}
      \H(X)& = \H(\vec{\beta})\\
      &\quad + \suml_{j=1}^d \E\left[N_j\right]\, \left(\H(\Exp(\lambda_j)) + \H(P_{1,j}, P_{2, j},\ldots,P_{d, j}, P_{d+1,j})\right),
    \end{aligned}
  \end{equation}
  which can be rewritten as
  \begin{align}
    \H(X) & = \H(\vec{\beta}) + \suml_{j=1}^d \E\left[N_j\right]\, \H(P_{1,j}, P_{2, j},\ldots,P_{d, j}, P_{d+1,j}) \label{eqn:H_discrete} \\
    &\quad + \suml_{j=1}^d \E\left[N_j\right]\, \H(\Exp(\lambda_j)).\label{eqn:H_continuous}      
  \end{align}
\end{myproposition}

\begin{proof}
  By virtue of Eq.~\eqref{eqn:H_occupation_time}, we replace $x^\ast_j/\vnorms{\vec{u}}$ by $\E\left[O_j\right]$ in Eq.~\eqref{eqn:H_X} and take into account that the entropy rate of a Poisson process with intensity rate $\lambda$ equals $\lambda\,(1-\log \lambda)$ to prove Eq.~\eqref{eqn:H_occupation_time}.
  To prove Eq.~\eqref{eqn:H_number_of_visits} we use Eq.~\eqref{eqn:N_i} to replace $x^\ast_j/\vnorms{\vec{u}}$ in Eq.~\eqref{eqn:H_X} by $\E\left[N_j\right]/\lambda_j$ and obtain
  \begin{equation}
    \begin{aligned}
      \H(X) &= -\suml_{i=1}^d\beta_i\,\log\beta_i + \suml_{j=1}^d \E\left[N_j\right](1-\log \lambda_j)\\
      &\qquad +\suml_{j=1}^d \E\left[N_j\right]\left(- \suml_{i=1,i\neq j}^d \frac{B_{ij}}{\lambda_j}\,\log \frac{B_{ij}}{\lambda_j} - \frac{z_j}{\lambda_j}\,\log \frac{z_j}{\lambda_j}\right).
    \end{aligned}
	\end{equation}
	Here, $(1-\log \lambda_j)$ is the entropy of an exponential random variable with rate parameter $\lambda_j$.
	Using definition \eqref{eqn:P_ij} of $P_{ij}$ we replace $B_{ij}/\lambda_j$ by $P_{ij}$ for $i\in S$ and $z_j/\lambda_j$ by $P_{d+1,j}$ and finish the proof.
\end{proof}

By identifying a compartmental system $M=M(\vec{u},\tens{B})$ with its associated absorbing continuous-time Markov chain $X$ and the according path $\mathcal{P}=\mathcal{P}(X)$ of a single traveling particle, we transfer the concept of the path entropy $\H(\mathcal{P})$ from the probabilistic to the deterministic realm.

\begin{definition}
  The \emph{path entropy of the compartmental system} $M$ in equilibrium given by Eq.~\eqref{eqn:lin_CS_sys} with associated absorbing continuous-time Markov chain $X$ and path $\mathcal{P}=\mathcal{P}(X)$, is defined by the path entropy 
\begin{equation}
  \H(\mathcal{P})=\H(\mathcal{P}(X))=\H(X).
\end{equation}
\end{definition}

Consider a one-dimensional compartmental system $M_\lambda$ in equilibrium with rate $\lambda>0$ and positive external input given by
\begin{equation}
  \deriv{t}\,x(t) = -\lambda\,x(t) + u,\quad t>0,
\end{equation}
and denote its associated path by $\mathcal{P}_\lambda$.
The entropy of the initial distribution vanishes, and we obtain
\begin{equation}
  \H(\mathcal{P}_\lambda) = \frac{x^\ast}{u}\,\lambda\,(1-\log\,\lambda) = \frac{1}{\lambda}\,\lambda\,(1-\log\,\lambda) = 1-\log\,\lambda,
\end{equation}  
which equals the differential entropy $1-\log\lambda$ of the exponentially distributed mean transit time $\TT_\lambda\sim\Exp(\lambda)$, reflecting that the only uncertainty of the particle's path in a one-pool system is the time of the particle's exit.
The exponential distribution with rate parameter $\lambda$ is the distribution of the interarrival time of a Poisson process wit intensity rate $\lambda$.
Hence, we can interpret $\H(\mathcal{P}_\lambda) = \lambda^{-1}\,\lambda\,(1-\log\lambda)$ as the instantaneous Poisson entropy rate $\lambda\,(1-\log\lambda)$ multiplied with the expected duration $\E\left[\TT_{\lambda}\right]=\lambda^{-1}$ of the particle's stay in the system.

Migrating to a $d$-dimensional system, we can interpret $\H(\mathcal{P})$ as the entropy of a continuous-time process in the light of Eq.~\eqref{eqn:H_occupation_time} and as the entropy of a discrete-time process in the light of Eq.~\eqref{eqn:H_number_of_visits}.
In both interpretations, the first term $\H(\vec{\beta})=\H(X_0)=\H(Y_1)$ represents the uncertainty of the first pool through which the particle enters the system.
In the continuous-time interpretation, the uncertainty of the subsequent travel is the weighted average of the superposition of $d$ Poisson processes describing the instantaneous uncertainty of possible jumps of the particle inside the system, $\theta(\Poi(B_{ij}))$, and out of the system, $\theta(\Poi(z_j))$, where the weights are the expected occupation times of the different compartments $j\in S$. 
In the discrete-time interpretation, the subsequent travel's uncertainty is the average of uncertainties associated to each pool, weighted by the number of visits to the respective pools.
The uncertainty associated to each pool comprises the uncertainty of the length of the stay in the pool, $\H(\Exp(\lambda_j))$, and the uncertainty of where to jump afterwards, $\H(\{P_{ij}:\,i\in\widetilde{S},\,j\in S,i\neq j\})$.
Hence, in the light of Eq.~\eqref{eqn:H_number_of_visits}, we can separate the path entropy into a discrete part associated to jump uncertainty given by Eq.~\eqref{eqn:H_discrete} and a continuous part associated to sojourn time uncertainty given by Eq.~\eqref{eqn:H_continuous}.

The two interpretations of the path entropy $\H(\mathcal{P})$ (as a continuous-time or discrete-time process) motivate two different entropy rates as described earlier.
The \emph{entropy rate per unit time} is given by
\begin{equation}
  \theta(\mathcal{P}) = \frac{\H(\mathcal{P})}{\E\left[\TT\right]}
\end{equation}
and the \emph{entropy rate per jump} by
\begin{equation}
  \theta_J(\mathcal{P}) = \frac{\H(\mathcal{P})}{\E\left[\mathcal{N}\right]}.
\end{equation}
While the path entropy measures the uncertainty of the entire path, entropy rates measure the average uncertainty of the instantaneous future of a particle while it is in the system: for the entropy rate per unit time the uncertainty entailed by the infinitesimal future, and for the entropy rate per jump the uncertainty entailed by the next jump. 
We can see that by considering the the stationary process
$Z=(Z_n)_{n\geq1}=(\widetilde{Y}_n,\widetilde{T}_n)_{n\geq1}$ on the space $(\widetilde{S}\times\R_+)$ defined by the transition probabilities $\widetilde{P}_{ij}(t) = \P(\widetilde{Y}_{n+1}=i, \widetilde{T}_{n+1}\leq t\,|\,\widetilde{Y}_n=j)$ given by
\begin{equation}
  \widetilde{P}_{ij}(t) = 
  \begin{cases}
     0, & i=j,\\
     B_{ij}\,\lambda_j^{-1}\,(1-e^{-\lambda_i\,t}), & i,j\leq d,\,i\neq j,\\
     z_j\,\lambda_j^{-1}, & i=d+1,\,j\leq d,\\
     \beta_i\,(1-e^{-\lambda_i\,t}), & i\leq d,\,j=d+1,
  \end{cases}
\end{equation}
and initial (stationary) distribution
\begin{equation}
  \pi_j(t) = \frac{1}{\E\left[\mathcal{N}\right]}\cdot
  \begin{cases}
    \E\left[N_j\right]\,(1-e^{-\lambda_j\,t}), & j\leq d,\\
    1, & j=d+1.
  \end{cases}
\end{equation}
This process describes the infinite journey, \ie\ the sequence of visited compartments with the associated sojourn times, of a single particle through the system with immediate jumps back into the system when leaving it.

\begin{myproposition}
  The entropy rate per jump, $\theta_J(\mathcal{P})$, equals the entropy rate of the stationary process $Z$.
\end{myproposition}

\begin{proof}
 Step 1. We show that $Z=(\widetilde{Y},\widetilde{T})$ is stationary.
 To that end, we define $\pi_j:=\lim_{t\to\infty} \pi_j(t)$, and we prove $\P(\widetilde{Y}_2=i,\widetilde{T}_2\leq t) = \pi_i(t) = \P(\widetilde{Y}_1=i,\widetilde{T}_1\leq t)$. Stationarity follows then by induction. Let $i=d+1$.
 Then,
\begin{equation}
  \begin{aligned}
    \P(\widetilde{Y}_2=i,\widetilde{T}_2\leq t) &= \suml_{j=1}^d \P(\widetilde{Y}_2=i,\widetilde{T}_2\leq t\,|\,\widetilde{Y}_1=j)\,\P(\widetilde{Y_1}=j)\\
    &= \suml_{j=1}^d \widetilde{P}_{d+1,j}(t)\,\pi_j\\
    &= \suml_{j=1}^d \frac{z_j}{\lambda_j}\,\frac{\E\left[N_j\right]}{\E\left[\mathcal{N}\right]}.
  \end{aligned}
\end{equation}
By Eq.~\eqref{eqn:N_i}, $r_j=z_j\,x^\ast_j$, and $\vnorms{\vec{r}}=\vnorms{\vec{u}}$, we get
\begin{equation}
    \P(\widetilde{Y}_2=i,\widetilde{T}_2\leq t)
    = \frac{1}{\E\left[\mathcal{N}\right]}\suml_{j=1}^d \frac{z_j}{\lambda_j}\,\frac{\lambda_j\,x^\ast_j}{\vnorms{\vec{u}}}
    = \frac{1}{\E\left[\mathcal{N}\right]}\,\frac{\vec{z}^T\,\vec{x}^\ast}{\vnorms{\vec{u}}} = \pi_{d+1}(t).
\end{equation}
Now let $i\leq d$.
Then
\begin{equation}
 \begin{aligned}
  \P(\widetilde{Y}_2=i,\widetilde{T_2}\leq t)
  &= \suml_{j=1,j\neq i}^d \frac{B_{ij}}{\lambda_j}\,(1-e^{-\lambda_i\,t})\,\frac{\E\left[N_j\right]}{\E\left[\mathcal{N}\right]} + \beta_i\,(1-e^{-\lambda_i\,t})\,\frac{1}{\E\left[\mathcal{N}\right]}\\
  &= \frac{1}{\E\left[\mathcal{N}\right]} \left[\suml_{i=1,i\neq j}^d \frac{B_{ij}\,x^\ast_j}{\vnorms{\vec{u}}} + \beta_i\right]\,(1-e^{-\lambda_i\,t})\\
  &= \frac{1}{\E\left[\mathcal{N}\right]} \left[\frac{(\tens{B}\,\vec{x}^\ast)_i}{\vnorms{\vec{u}}} - \frac{B_{ii}\,x^\ast_i}{\vnorms{\vec{u}}} + \beta_i\right]\,(1-e^{-\lambda_i\,t})\\
  &= \frac{1}{\E\left[\mathcal{N}\right]}\left[-\frac{u_i}{\vnorms{\vec{u}}} + \frac{\lambda_i\,x^\ast_i}{\vnorms{\vec{u}}} + \beta_i\right]\,(1-e^{\lambda_i\,t})\\
  &= \frac{1}{\E\left[\mathcal{N}\right]}\,\E\left[N_i\right]\,(1-e^{-\lambda_i\,t})\\
  &= \pi_i(t).
 \end{aligned}
\end{equation} 

Step 2. Since $Z$ is stationary, by \citet[Theorem~4.2.1]{Cover2006}, its entropy rate given by
\begin{equation}
  \theta(Z) = \liml_{n\to\infty} \H(Z_{n+1}\,|\,Z_n) = \H(Z_2\,|\,Z_1),
\end{equation}
which computes to
\begin{equation}
 \begin{aligned}
   \theta(Z) 
   &= \H((\widetilde{Y}_2,\widetilde{T}_2)\,|\,(\widetilde{Y}_1,\widetilde{T}_1))
   = \H((\widetilde{Y}_2,\widetilde{T}_2)\,|\,\widetilde{Y}_1)
   = \H(\widetilde{T}_2\,|\,\widetilde{Y}_2,\widetilde{Y}_1) + \H(\widetilde{Y}_2\,|\,\widetilde{Y}_1)\\
   &= \H(\widetilde{T}_2\,|\,\widetilde{Y}_2) + \H(\widetilde{Y}_2\,|\,\widetilde{Y}_1).
 \end{aligned}
\end{equation}
By stationarity, $\H(\widetilde{T}_2\,|\,\widetilde{Y}_2) = \H(\widetilde{T}_1\,|\,\widetilde{Y}_1)$.
Consequently,
\begin{equation}
 \begin{aligned}
  \theta(Z)
  &= \H(\widetilde{T}_1\,|\,\widetilde{Y}_1) + \H(\widetilde{Y}_2\,|\,\widetilde{Y}_1)\\
  &= \suml_{j=1}^{d+1} \pi_j\,\left[\H(\widetilde{T_1}\,|\,\widetilde{Y}_1=j) + \H(\widetilde{Y}_2\,|\,\widetilde{Y}_1=j)\right],\\
  &= \frac{1}{\E\left[\mathcal{N}\right]} \left(\suml_{j=1}^d \E\left[N_j\right] \left[\H(\widetilde{T}_1\,|\,\widetilde{Y}_1=j) + \H(\widetilde{Y}_2\,|\,\widetilde{Y}_1=j)\right] + \H(\widetilde{Y}_2\,|\,\widetilde{Y}_1=d+1)\right),
 \end{aligned}
\end{equation}
which together with Eq.~\eqref{eqn:H_number_of_visits} finishes the proof.
\end{proof}
If we divide the entropy rate per jump by the average time between two jumps, we obtain the entropy rate per unit time.
The average time between two jumps is given by
\begin{equation}
  \suml_{j=1}^d\pi_j\,\lambda_j^{-1} = \frac{1}{\E\left[\mathcal{N}\right]}\,\suml_{j=1}^d \frac{x^\ast_j}{\vnorms{\vec{u}}} = \frac{\E\left[\TT\right]}{\E\left[\mathcal{N}\right]}.
\end{equation}
Hence,
\begin{equation}
  \theta(\mathcal{P}) = \frac{\E\left[\mathcal{N}\right]}{\E\left[\TT\right]}\,\theta(Z)
\end{equation}
is the average uncertainty per unit time of the stationary process $Z$.

\subsection{The maximum entropy principle (MaxEnt)}
MaxEnt arose in statistical mechanics as a variational principle to predict the equilibrium states of thermal systems and later was applied to matters of information and as a general procedure to draw inferences based on self-consistency requirements \citep{Presse2013RMP}.
Its relationship to information theory and stochastics was established by \citet{Jaynes1957PR1, Jaynes1957PR2}.
The general idea is to identify the most uninformed probability distribution to represent some given data in the sense that the maximum entropy distribution, constrained to given data, uses the information provided by the data only and nothing else.
This approach ensures that no additional subjective information creeps into the distribution.
The goal of this section is to transfer MaxEnt to compartmental systems in order to identify the compartmental system that represents our state of knowledge best in different situations, and at the same time get a better understanding of the introduced entropy measures.
In the next two examples we identify compartmental models with maximum entropy under some restrictions.
Both examples show that maximizing entropy means also maximizing symmetry, as much as the given constraints allow.

\begin{example}
\label{max_ent_example_1}
Consider the set $\mathcal{M}_1$ of equilibrium compartmental systems~\eqref{eqn:lin_CS_sys} with a predefined nonzero input vector $\vec{u}$, a predefined mean transit time $\E\left[\TT\right]$, and an unknown steady-state vector $\vec{x}^\ast$ comprising nonzero components.
We are interested in the most unbiased compartmental system that reflects our state of information, where maximum unbiasedness is achieved by identifying $M^\ast_1\in\mathcal{M}_1$ with path $\mathcal{P}^\ast_1:=\mathcal{P}(M^\ast_1)$ such that the path entropy $\H(\mathcal{P}^\ast_1)$, or equivalently, the entropy rate per unit time $\theta(\mathcal{P}^\ast_1)$ is maximized. 
We can show (see Proposition~\ref{proposition:max_ent_example_1}) that the compartmental system $M^\ast_1=M(\vec{u},\tens{B})$ with 
\begin{equation}
	\tens{B} = \begin{pmatrix}
    -\lambda & 1 & \cdots & 1\\
		1 & -\lambda & 1 \cdots & 1 \\
		\vdots & & \ddots & \vdots\\
		1 & \cdots & 1 & -\lambda
  \end{pmatrix},
\end{equation}
where $\lambda=d-1+1/\E\left[\TT\right]$, 		
is the maximum entropy model in $\mathcal{M}_1$.
In the special case $d=1$ for a one-dimensional compartmental system, we obtain $B=-1/\E\left[\TT\right]$.
Since in this case $\TT\sim\Exp(-B)$, we see that the exponential distribution is the maximum entropy distribution in the class of all nonnegative continuous probability distributions with fixed expected value.
This special case is very well known \citep[Example~12.2.5]{Cover2006}.
\end{example}

\begin{example}
\label{max_ent_example_2}
Let us consider the subclass $\mathcal{M}_2\subseteq\mathcal{M}_1$ of compartmental models from the previous example with the additional restriction of a predefined positive steady-state vector $\vec{x}^\ast$.
Then the compartmental system $M^\ast_2=M(\vec{u},\tens{B})$ with path $\mathcal{P}^\ast_2$ and
\begin{equation}
	B_{ij} = \begin{cases}
    \sqrt\frac{x_i^\ast}{x_j^\ast},\quad & i\neq j,\\
		-\suml_{k=1,k\neq j}^d \sqrt\frac{x_k^\ast}{x_j^\ast} - \frac{1}{\sqrt{x_j^\ast}}, \quad &i=j,
		\end{cases}
  \end{equation}
is the maximum entropy model in $\mathcal{M}_2$ (see Proposition~\ref{proposition:max_ent_example_2}).
\end{example}

\subsection{Structural model identification assisted by MaxEnt}
Suppose we observe a natural system and conduct measurements from which we try to construct a linear autonomous compartmental model in equilibrium that represents the observed natural system as well as possible.
The first question that arises is the one for the number of compartments the model should ideally have.
MaxEnt cannot be helpful here because by adding more and more compartments we can theoretically increase the entropy of the model indefinitely.
Consequently, the problem of finding the right dimension of system~\eqref{eqn:lin_CS_sys} has to be solved by other means.
One way to do this is to analyze an impulse response function of the system and its Laplace transform, \ie\ the transfer function of the system, and identify the most dominating frequencies.
The impulse response or the transfer function might be possible to obtain by tracer experiments \citep{Anderson1983, Walter1986MBS}.

Once the desired number of compartments is identified, we can focus on the structure and values of external input and output fluxes as well as internal fluxes.
In \citet[Chapter~16]{Anderson1983} the \emph{structural identification problem} of linear autonomous systems is described as follows.
Suppose we are interested in determining a $d$-dimensional system of form~\eqref{eqn:lin_CS_sys}.
We are interested in sending an impulse into the system at time $t=0$ and analyzing its further behavior.
To that end, we rewrite the system to
\begin{equation}\label{eqn:ABC_system}
	\begin{aligned}
		\deriv{t}\,\vec{x}(t) &= \tens{B}\,\vec{x}(t) + \tens{A}\,\vec{u},	\quad t\geq0,\\
		\vec{x}(0) &= \vec{0}\in\R^d,\\
		\vec{y}(t) &= \tens{C}\,\vec{x}(t),\quad t\geq0.
	\end{aligned}
\end{equation}
Note that the roles of $\tens{A}$ and $\tens{B}$ are interchanged here with respect to \citet{Anderson1983}.
In a typical tracer experiment, we choose an input vector $\vec{u}$ and the \emph{input distribution matrix} $\tens{A}$, which defines how the input vector enters the system.
Then we decide which compartments we can observe to determine the \emph{output connection matrix} $\tens{C}$.
The experiment is now to inject an impulse into the system and to record the output function $\vec{y}(t) =\tens{C}\,\vec{x}(t)$.
\citet{Bellman1970MBS} pointed out that the input-output relation is given by
\begin{align*}
	\vec{y}(t) &= \tens{C}\,\vec{x}(t) = \tens{C}\,\intl_0^t e^{(t-\tau)\,\tens{B}}\,\tens{A}\,\vec{u}(\tau)\dd{\tau}\\
	&= \left[\tens{C}\,e^{t\,\tens{B}}\,\tens{A}\right] * \vec{u}(t),
\end{align*}
where $*$ is the convolution operator.
The model parameters enter the input-output relation only in the matrix-valued \emph{impulse response function}
\begin{equation}
	\tens{\Psi}(t):=\tens{C}\,e^{t\,\tens{B}}\,\tens{A},\quad t\geq0,
\end{equation}
or in the \emph{transfer function}
\begin{equation}
	\widehat{\tens{\Psi}}(s) := \tens{C}\,(s\,\tens{I}-\tens{B})^{-1}\,\tens{A}, \quad s\geq0,
\end{equation}
which is the Laplace transform matrix of $\tens{\Psi}$.
Consequently, all identifiable parameters of $\tens{A}$, $\tens{B}$, and $\tens{C}$ must be identified through $\tens{\Psi}$ or $\widehat{\tens{\Psi}}$.
Difficulties arise because the entries of the matrices $\tens{\Psi}$ and $\widehat{\tens{\Psi}}$ are usually nonlinear expressions of the elements of $\tens{A}$, $\tens{B}$, and $\tens{C}$.
We call system~\eqref{eqn:ABC_system} \emph{identifiable} if this nonlinear system of equations has a unique solution $(\tens{A},\tens{B},\tens{C})$ for given $\tens{\Psi}$ or $\widehat{\tens{\Psi}}$.
Otherwise the system is called \emph{non-identifiable}.
Usually, the matrices $\tens{A}$ and $\tens{C}$ are already know from the experiment's setup.
What remains is to identify the compartmental matrix $\tens{B}$, and this can be done by MaxEnt.

\section{Application to particular systems}
First, we apply the presented theory to some equilibrium compartmental models with very simple structure in order to get some grasp on the new entropy concepts.
Then we compute entropy quantities for two carbon-cycle models in dependence on environmental and biochemical parameters.
At last we apply MaxEnt to solve an equifinality problem in model selection as an example for how to tackle this problem arising from, for instance, tracer experiments.

\subsection{Simple examples}
\label{sec:simple_examples}
From Table~\ref{tab:entropy_table} we can see that, depending on the connections between compartments, smaller systems can have greater path entropy and entropy rates than bigger systems, even though systems with more compartments can theoretically reach higher entropy.
Furthermore, we see from the depicted examples that the system with the highest path entropy does neither have the highest entropy rate per unit time nor per jump.
Adding connections to a system, one would expect higher path entropy, but the path entropy might actually decrease because the new connections potentially provide a faster way out the system.

\begin{landscape}
\begin{table}[htbp]
   \centering
   \caption{Overview of different entropy measures of simple models with different structures. The columns from left to right represent a schematic representation of the model structure, its mathematical representation, entropy rate per jump $\theta_J$, mean number of jumps $\mathbb{E} [\mathcal{N}]$, entropy rate per unit time $\theta$, mean transit time $\mathbb{E} [\mathcal{T}]$, and path entropy $\H(\mathcal{P})$. Underlined numbers are the highest values per column }
   \begin{tblr}{
     colspec = {X[c,h,4cm]X[c,3.5cm]X[c, 2.5cm]X[c,1cm]X[c,2.5cm]X[c,1cm]X[c,1cm]}
    }
        \hline
        Structure    & $\frac{\mathrm{d}}{\mathrm{d}t} \mathbf{x} (t)$ & $\theta_J$ & $\mathbb{E} [\mathcal{N}]$ & $\theta$ & $\mathbb{E} [\mathcal{T}]$ & $\H(\mathcal{P})$ \\
        \hline
         \input{figs/Compartments/onepool.tex} & $-\lambda x +1$ & $0.5 \ (1 - \log \lambda)$ & 2.00 & $\lambda (1 - \log \lambda)$ & $1/ \lambda$ & $1 - \log$ \\
         \input{figs/Compartments/twoseries.tex}    & $\left( \begin{matrix} -1 & 0 \\ 1 & -1  \end{matrix} \right) x + \left( \begin{matrix} 1 \\ 0 \end{matrix} \right)$ & 0.67 & 3.00 & 1.00 & 2.00 & 2.00  \\
         \input{figs/Compartments/twoparallel.tex}  & $\left( \begin{matrix} -1 & 0 \\ 0 & -1  \end{matrix} \right) x + \left( \begin{matrix} 1 \\ 1 \end{matrix} \right)$ & 0.85 & 2.00 & 1.69 & 1.00 & 1.69 \\
         \input{figs/Compartments/twofeedback1.tex}  & $\left( \begin{matrix} -1 & 1/2 \\ 1 & -1  \end{matrix} \right) x + \left( \begin{matrix} 1 \\ 0 \end{matrix} \right)$ & 1.08 & 5.00 & 1.35 & \underline{4.00} & \underline{5.39} \\
         \input{figs/Compartments/twofeedback2.tex}   & $\left( \begin{matrix} -1 & 1/2 \\ 1/2 & -1  \end{matrix} \right) x + \left( \begin{matrix} 1 \\ 1 \end{matrix} \right)$ & \underline{1.36} & 3.00 & 2.04 & 2.00 & 4.08 \\
         \input{figs/Compartments/threeseries.tex}   & $\left( \begin{matrix} -1 & 0 & 0 \\ 1 & -1 & 0 \\ 0 & 1 & -1 \end{matrix} \right) x + \left( \begin{matrix} 1 \\ 0 \\ 0 \end{matrix} \right)$ & 0.75 & 4.00 & 1.00 & 3.00 & 3.00 \\
         \input{figs/Compartments/threeparallel.tex}   & $\left( \begin{matrix} -1 & 0 & 0 \\ 0 & -1 & 0 \\ 0 & 0 & -1 \end{matrix} \right) x + \left( \begin{matrix} 1 \\ 1 \\ 1 \end{matrix} \right)$ & 1.05 & 2.00 & \underline{2.10} & 1.00 & 2.10 \\
     \hline
   \end{tblr}
   \label{tab:entropy_table}
\end{table}

\end{landscape}

\subsection{A linear autonomous global carbon-cycle model}
\label{sec:example_1}
\begin{figure}[htbp]
  \centering
  \includegraphics[width=\linewidth]{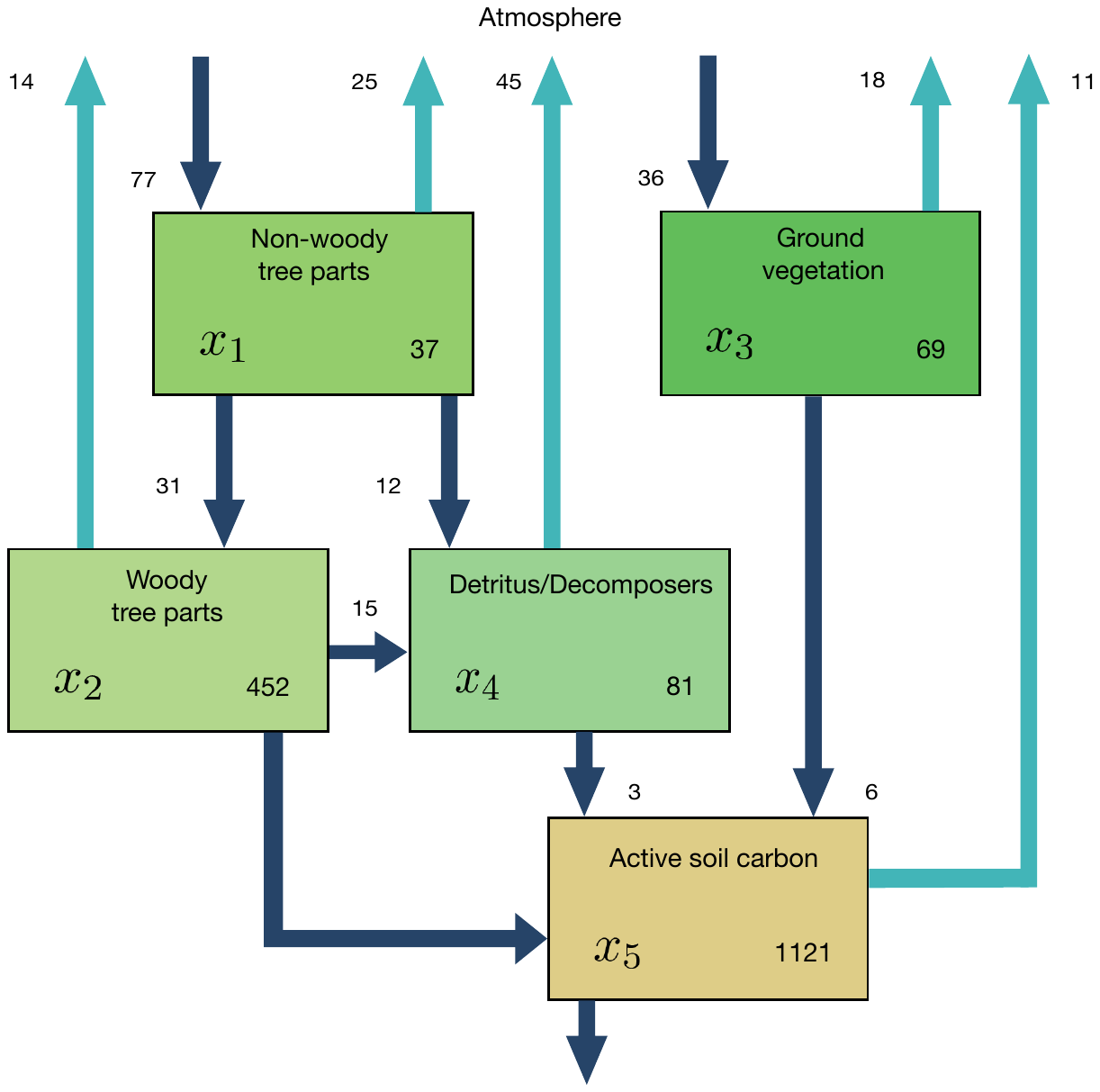}
  \caption{Schematic of the linear autonomous global carbon cycle model in steady state introduced by \citet{Emanuel1981}
  }
  \label{fig:Emanuel_model}
\end{figure}

\begin{figure}[htbp]
    \centering
    \includegraphics[width=1.0\linewidth]{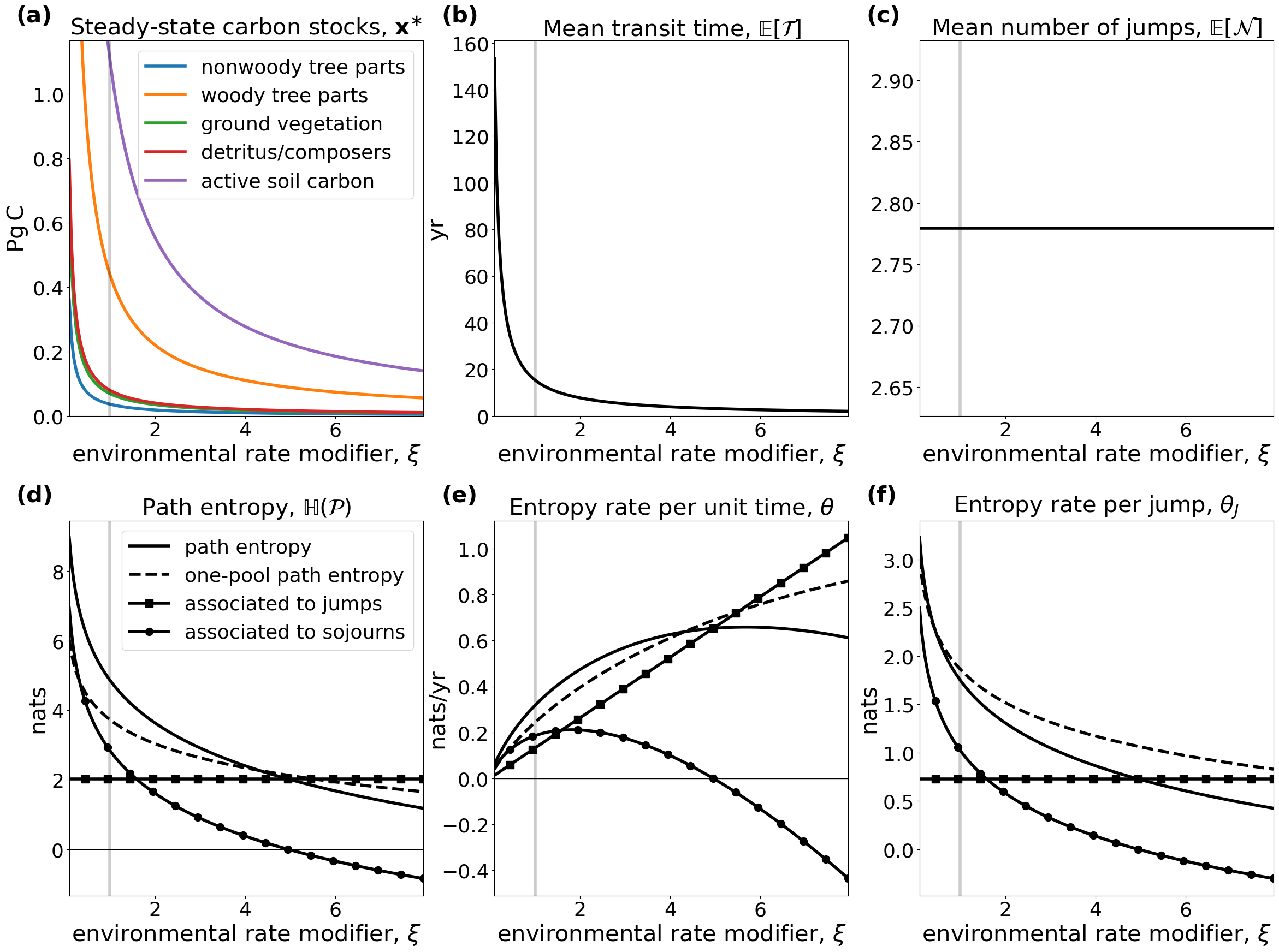}
    \caption{
    (a) Equilibrium carbon stocks and (b)--(f) entropy related quantities of the global carbon cycle model introduced by \citet{Emanuel1981} in dependence on the environmental rate coefficient $\xi$
  }
  \label{fig:Emanuel_entropies}
\end{figure}

We consider the global carbon-cycle model introduced by \citet{Emanuel1981} (Fig.~\ref{fig:Emanuel_model}).
The model comprises five compartments: non-woody tree parts $x_1=37\,\peta\gC$, woody tree parts $x_2=452\,\peta\gC$, ground vegetation $x_3=69\,\peta\gC$, detritus/decomposers $x_4=81\,\peta\gC$, and active soil carbon $x_5=1,121\,\peta\gC$.
We introduce an environmental rate modifier $\xi$ which controls the speed of the system.
This parameter could potentially increase and speed up the system with increasing global surface temperature \citep{Sierra2023PTRS}.
For a given $\xi$ the equilibrium model $M_\xi=M(\vec{u},\,\tens{B}_\xi)$ is given by
\begin{equation}
  \vec{u} = (77;\,0;\,36;\,0;\,0)^{\transpose}\, \peta\gC\,\yr^{-1}
\end{equation}
and
\begin{equation}
    \tens{B}_\xi = \xi\,\left(\begin{matrix}
      -77/37 &       0 &      0 &      0 & 	  0\\
       31/37 & -31/452 &      0 &      0 & 	  0\\
	   0 &       0 & -36/69 &      0 & 	  0\\
       21/37 &  15/452 &  12/69 & -48/81 & 	  0\\
	   0 &   2/452 &   6/69 &   3/81 & -11/1,121
	 \end{matrix}\right)\,\yr^{-1},
\end{equation}
where the numbers are chosen as in \citet{Thompson1999GCB}. 
The input vector is expressed in units of petagrams of carbon per year ($\peta\gC\,\yr^{-1}$) and the fractional transfer coefficients in units of per year ($\yr^{-1}$).
Because $\tens{B}_\xi$ is a lower triangular matrix, the model contains no feedbacks.
For every value of $\xi$ the system has a different steady state (Fig.~\ref{fig:Emanuel_entropies}, panel (a)).
The higher the value of $\xi$, the faster the system, which makes the mean transit time (panel (b)) decrease, and because of shorter paths also the path entropy (panel (d)) decreases.
Since $\xi$ has no impact on the structure of the model, the mean number of jumps (panel (c)) remains unaffected.
This can also be seen from the solid line marked by squares in panel (d).
It represents the part of the path entropy related to jump-associated uncertainties (Eq.~\eqref{eqn:H_discrete}).
The solid line marked by circles represents the part of the path entropy related to sojourn-associated uncertainties (Eq.~\eqref{eqn:H_continuous}), which as a weighted average of one-pool entropies decreases similarly as the entropy of an exponential distribution with increasing rate parameter $\lambda$ (Fig.~\ref{fig:simple_entropy}, panel (b)).
The two parts together constitute the path entropy as represented by the unmarked solid line.

The entropy rate per unit time (panel (e)) increases until $\xi\approx6$ and decreases afterwards, because with increasing system speed the decreasing uncertainty associated to sojourn times increasingly dominates the uncertainty associated to jumps.
While the uncertainty associated to jumps averaged over the path length increases because the total jump uncertainty is constant (see solid line marked with squares in panel (d)) and the mean path length decreases (panel (b)), the sojourn-associated uncertainty decreases with increasing system speed for $\xi>6$ similar to the entropy rate of a Poisson process with intensity rate $\lambda>1$ (see Fig.~\ref{fig:simple_entropy}, panel (c)).
The entropy rate per jump (panel (f)) decreases with increasing $\xi$, because the path entropy of the system decreases.

Dashed lines in panels (d)--(f) show the respective entropy values for a one-pool system $M_\lambda=M((77+36)\,\peta\gC\,\yr^{-1},\, -\lambda)$ with the same mean transit time, \ie\ $\lambda^{-1} = \E\left[\TT_\xi\right]$.
The solid and dashed lines intersect at $\xi\approx4.31$ in panels (d) and (e).
Before this break-even point the path of this multiple-pool model is harder to predict than the path (\ie\ the exit time of the particle) of a one-pool model with the same mean transit time.
After this point of break even, the path of the model with five compartments is easier to predict than only the transit time in a one-pool model.
The reason is that as the system becomes faster, the differential entropy of the sojourn times in slow pools decreases so fast that at some point the sojourn times in slow pools visited by few particles becomes rather unimportant.
The one-pool model's path becomes relatively harder to predict because it puts too much weight on a small amount of slowly cycling particles.

Note that there is no point in comparing jump-associated uncertainties (square-marked lines) with one-pool entropies (dashed lines), because the former are discrete entropies and the latter differential entropies.
Comparison of a differential entropy with another quantity becomes only reasonable if a second differential entropy is involved as is true for the path entropy or the entropy rates $\theta$ and $\theta_J$ (unmarked solid lines).
Hence, square- and circle-marked lines assist in understanding the composition of the entropies of the multi-pool system, and only the composition of the two can then be compared to the one-pool entropy rate.

\subsection{A nonlinear autonomous soil organic matter decomposition model}
\label{sec:example_2}
\begin{figure}[htbp]
    \centering
    \includegraphics[width=0.8\linewidth]{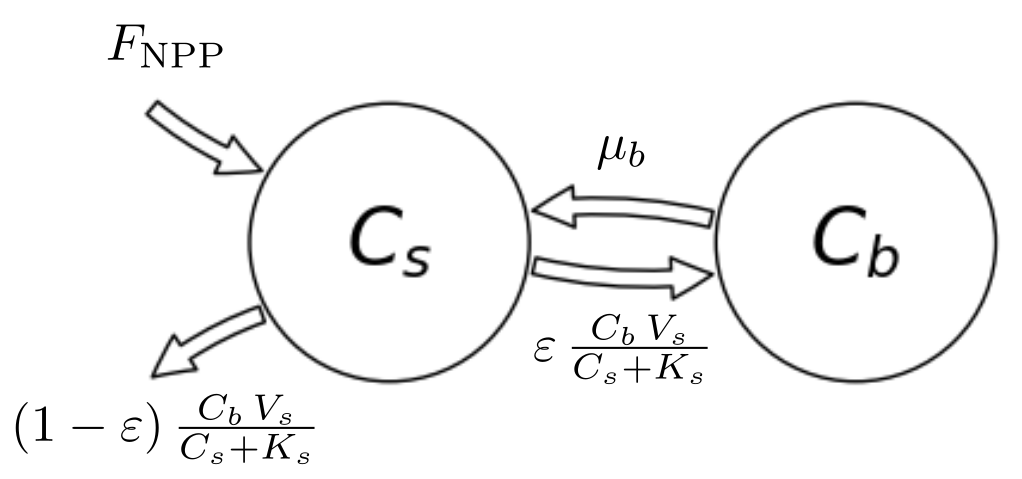}
    \caption{
    Scheme of the nonlinear autonomous carbon cycle model introduced by \citet{Wang2014BG} with two compartments: substrate organic carbon ($C_s$) and microbial biomass ($C_b$)
    }
    \label{fig:Wang_model}
\end{figure}

Consider the nonlinear two-compartment model $M_\varepsilon=M(\vec{u},\,\tens{B}_\varepsilon)$ described by \citet{Wang2014BG} which is used to represent the dynamics of microbes and carbon substrates in soils (Fig.~\ref{fig:Wang_model}). Its ODE system is given by
\begin{equation}
    \deriv{t}\,\begin{pmatrix}C_{s}\\C_{b}\end{pmatrix}(t) = 
    \begin{pmatrix}
      -\lambda(\vec{x}(t)) & \mu_{b}\\
      \varepsilon \lambda(\vec{x}(t)) & - \mu_{b}
    \end{pmatrix}
    \, \begin{pmatrix}C_{s}\\C_{b}\end{pmatrix}
    + \begin{pmatrix}F_{\NPP}\\0\end{pmatrix},
\end{equation}
where $\vec{x}(t)=(C_{s},\,C_{b})^{\transpose}(t)$.
We denote by $C_s$ and $C_b$ substrate organic carbon and soil microbial biomass carbon ($\gC\,\meter^{-2}$), respectively, by $\varepsilon$ the carbon use efficiency or fraction of assimilated carbon that is converted into microbial biomass (unit-less), by $\mu_b$ the turnover rate of microbial biomass per year ($\yr^{-1}$), by $F_{\NPP}$ the carbon influx into the soil ($\gC\,\meter^{-2}\,\yr^{-1}$), and by $V_s$ and $K_s$ the maximum rate of soil carbon assimilation per unit microbial biomass per year ($\yr^{-1}$) and the half-saturation constant for soil carbon assimilation by microbial biomass ($\gC\,\meter^{-2}$), respectively.

We consider the model in equilibrium, \ie\ $\vec{x}(t)=\vec{x}^\ast=(C_s^\ast,\,C_b^\ast)^T$ with
\begin{equation}
  C_s^\ast = \frac{K_{s}}{\frac{V_{s} \varepsilon}{\mu_{b}} - 1}\quad\text{ and }\quad C_b^\ast = \frac{F_{\NPP}}{\mu_{b} \left(-1 + \frac{1}{\varepsilon}\right)}.
\end{equation}
The equilibrium stocks depend on the carbon use efficiency $\varepsilon$ and so does the compartmental matrix $\tens{B}=\tens{B}_\varepsilon$, because
\begin{equation}\label{eqn:lambdax}
    \lambda(\vec{x}) = \frac{C_{b} V_{s}}{C_{s} + K_{s}}.
\end{equation}
From \citet{Wang2014BG} we take the parameter values $\mu_b = 4.38\,\yr^{-1}$, $F_{\NPP} = 345.00\,\gC\,\meter^{-2}\,\yr^{-1}$, and $K_s = 53,954.83\,\gC\,\meter^{-2}$.
Since the description of $V_s$ is missing in the original publication, we let it be equal to $59.13\,\yr^{-1}$ to approximately meet the given steady-state contents $C_s^\ast = 12,650.00\,\gC\,\meter^{-2}$ and $C_b^\ast = 50.36\,\gC\,\meter^{-2}$ for the original value $\varepsilon=0.39$.
Otherwise we leave the carbon use efficiency $\varepsilon$ as a free parameter.

\begin{figure}[htbp]
    \centering
    \includegraphics[width=1.0\linewidth]{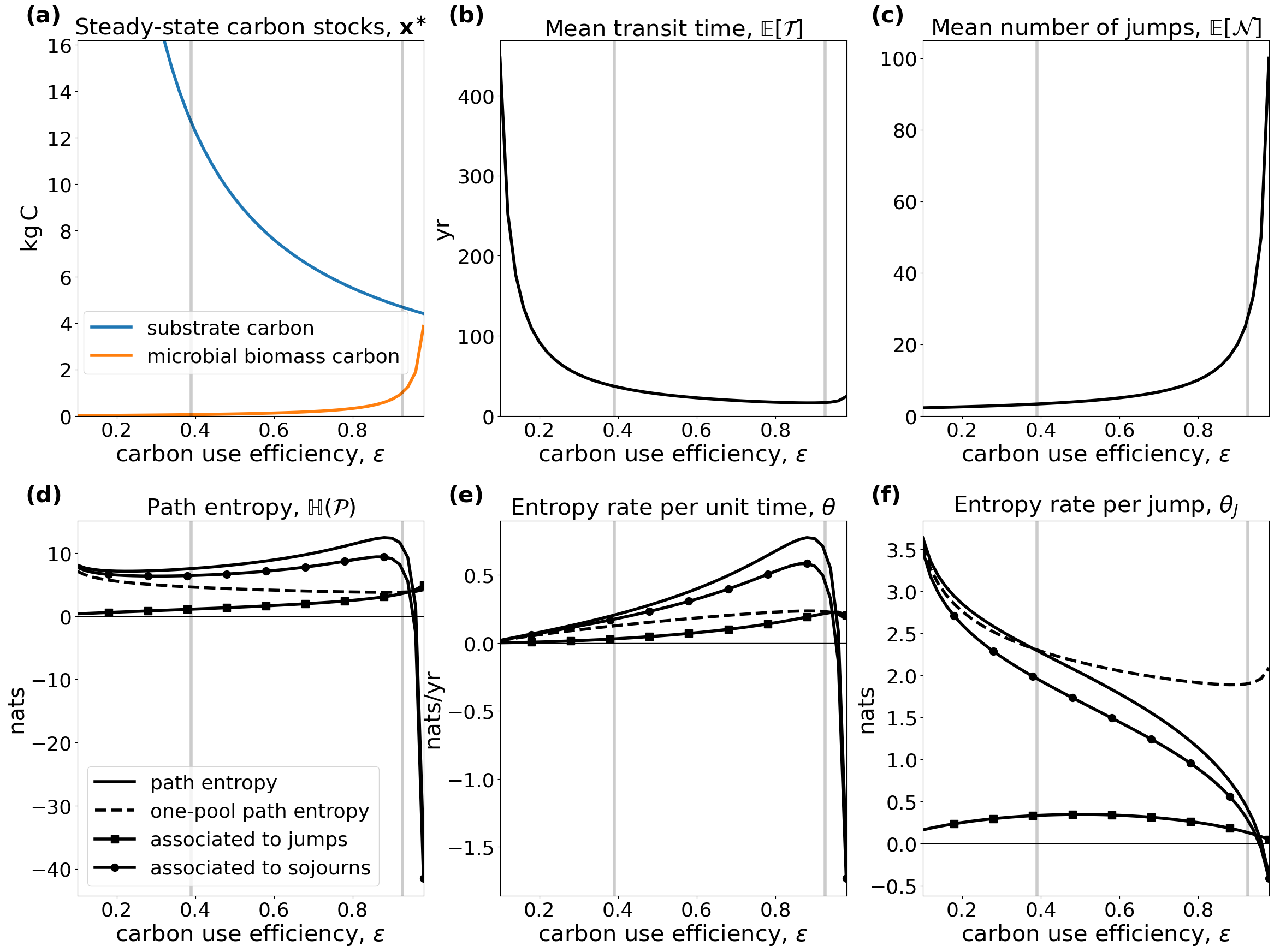}
    \caption{
    (a) Equilibrium carbon stocks and (b)--(f) entropy related quantities of the global carbon cycle model introduced by \citet{Wang2014BG} in dependence on the microbial carbon use efficiency $\varepsilon$
    }
    \label{fig:Wang_entropies}
\end{figure}

In contrast to the system from the first example, this system exhibits a feedback.
This feedback results from dead soil microbial biomass being considered as new soil organic matter.
The feedback can also be recognized by noting that $\tens{B}$ is not triangular.
For every value of $\varepsilon$ the system has a different steady state (Fig.~\ref{fig:Wang_entropies}, panel (a)).
The higher the value of $\varepsilon$, the lower the equilibrium substrate organic carbon and the higher the microbial biomass carbon.
Caused by the model's nonlinearity expressed in Eq.~\eqref{eqn:lambdax}, the system speed increases and the mean transit time goes down (panel (b)) with increasing $\varepsilon$.
At the same time, higher carbon use efficiency increases the probability of each carbon atom to be reused more often, hence the mean number of jumps increases (panel (c)), making the entropy rate per jump decrease (panel (f)).
Even though the average paths become shorter, with increasing carbon use efficiency the path entropy increases as well for most values of $\varepsilon$.
This has two reasons.
First, the mean uncertainty of where to jump from $C_s$ increases, this uncertainty decreases then for $\varepsilon>0.5$ (solid line marked by squares in panel (f)).
Second, the rate $-B_{11}$ of leaving the substrate pool is increasing and smaller than $1$.
The corresponding Poisson process reaches its maximum entropy rate at an intensity rate equal to $1$ (Fig.~\ref{fig:simple_entropy}, panel (c)), which corresponds to $\varepsilon\approx0.926$.
This is also reflected in the entropy rate per unit time (panel (e)).
The maximum does not exactly occur at $\varepsilon=0.926$, because the times that the particle stays in the different pools also depends on $\varepsilon$.
For $\varepsilon$ approaching $1$, both the path entropy and the entropy rate rapidly decline as the sojourn-associated uncertainties (solid lines with circle markers) decline sharply because of a nonlinear increase of the rate $-B_{11}$ of soil organic carbon turnover.

Considering a one-pool system $M_\lambda=M(345.00\,\gC\,\meter^{-2}\,\yr^{-1},\, -1/\E\left[\TT_\varepsilon\right])$ with the same mean transit time, we recognize only small sensitivity of the entropies on $\varepsilon$, because the contrary effects on path length and jump- and sojourn-associated uncertainties mostly balance out (dashed lines in panels (d)--(f)).

\subsection{Model identification via Maxent}
  \label{sec:model_identification}
  The following example is inspired by \citet[Example~16\,C]{Anderson1983}.
  It shows how MaxEnt can help make a decision about which model to use if not all parameters can be uniquely determined from the transfer function $\widehat{\tens{\Psi}}$.
	We are interested in determining the entries of the compartmental matrix $\tens{B}$ belonging to the $2$-dimensional equilibrium compartmental system
	\begin{equation}\label{eqn:opt_example}
        \deriv{t}\,
		\begin{pmatrix} x_1 \\ x_2 \end{pmatrix}(t)
		=
		\begin{pmatrix} B_{11} & B_{12} \\ B_{21} & B_{22} \end{pmatrix}\,
		\begin{pmatrix} x_1 \\ x_2 \end{pmatrix}(t)
		+
		\begin{pmatrix} 1 \\ 0 \end{pmatrix}\,\gC\,\yr^{-1},
		\quad t>0.
	\end{equation}
	We immediately notice that $\vec{u}=(1,0)^T\,\gC\,\yr^{-1}$ and $\tens{A}=\tens{I}$.
	Further, we decide to measure the contents of compartment $1$ such that $\tens{C} = (1,0)$.
	We recall $z_j = -\sum_{i=1}^d B_{ij}$ and obtain $z_1 = -B_{11} - B_{21}$ and $z_2 = - B_{22} - B_{12}$.
	The real-valued transfer function is then given by
	\begin{equation}
		\widehat{\tens{\Psi}}(s) = \frac{s + \gamma_1}{s^2+\gamma_2\,s+\gamma_3},
	\end{equation}
	where
	\begin{equation}\label{eqn:measurement_data}
		\begin{aligned}
			\gamma_1 &= B_{12} + z_2,\\
			\gamma_2 &= B_{21} + z_1 + B_{12} + z_2,\\
			\gamma_3 &= z_1\,B_{12} + z_1\,z_2 + B_{21}\,z_2.
		\end{aligned}
	\end{equation}
	We assume that $\widehat{\tens{\Psi}}$ is known from measurements, \ie, $\gamma_1$, $\gamma_2$, and $\gamma_3$ are known impulse response parameters.	
	We have the four unknown parameters $B_{11}$, $B_{12}$, $B_{21}$, and $B_{22}$, or equivalently, $B_{12}$, $B_{21}$, $z_1$, and $z_2$, but only three equations to determine them.
	Consequently, the system is non-identifiable and there remains a class $\mathcal{M}$ of models which all satisfy Eq.~\eqref{eqn:measurement_data}.
	Which model out of $\mathcal{M}$ are we going to select now?

	Here, MaxEnt comes into play.
	We intend to select the model that best represents the information given by our measurement data.
	We have to find $M^\ast=M(\vec{u},\tens{B}^\ast)$ such that
	\begin{equation}
		M^\ast = \operatornamewithlimits{arg\,max}\limits_{M\in\mathcal{M}}\,\theta(\mathcal{P}(M)).
	\end{equation}
	Maximizing the entropy rate per unit time here leads to a feasible optimization problem, whereas maximization of the path entropy by slowing down the model and indefinitely increasing its mean transit time and with it its path entropy would lead to an unbounded optimization problem.
	The parameter space associated to $\mathcal{M}$ is given by
	\begin{equation}
	 \{\vec{p}=(B_{12}, B_{21}, z_1, z_2)\in\R^4_+:\,\vec{p}\text{ satisfies Eq.~\eqref{eqn:measurement_data}}\},
	\end{equation}
	and is not guaranteed to be convex in general.
	Consequently, by fundamental principles from mathematical optimization theory, existence and uniqueness of $M^\ast$ are not guaranteed and we must apply optimization methods tailored to the specific case at hand.

\begin{figure}[ht]
    \centering
    \includegraphics[width=0.95\linewidth]{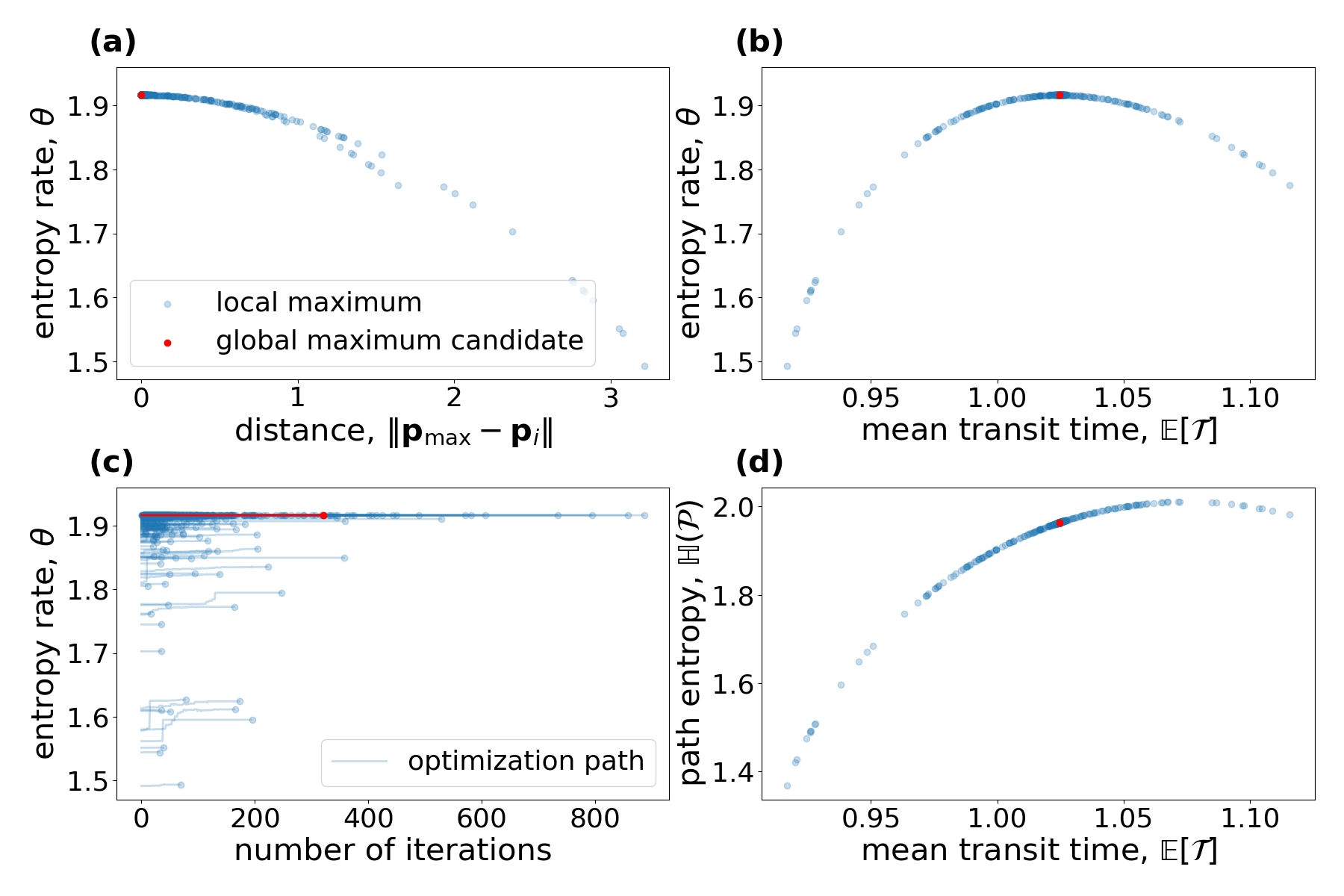}
	\caption{
		Local maximizations of $\theta$ over a grid on a subspace of the parameter space. For better visibility we chose randomly $1,000$ grid points for the plot. Blue dots show local maxima found during the global maximization procedure starting on the grid. 
		The red dot is associated to the global maximum candidate $M_{\text{max}}$.
		(a) Entropy rate per unit time versus $l_1$-distance of local maxima $\vec{p}_i$ parameters from the global maximum candidate parameters $\vec{p}_{\text{max}}$.
		(b) Entropy rate per unit time versus mean transit time.
		(c) Paths of entropy rate per unit time during the local maximizations on the grid.
		(d) Path entropy versus mean transit time 
    }
		\label{fig:model_id}
    \end{figure}

	Let us turn to a numerical example in which we suppose to be given $\gamma_1=3\,\yr^{-1}$, $\gamma_2=5\,\yr^{-1}$, and $\gamma_3=4\,\yr^{-1}$.
	Since convexity of the parameter space is not guaranteed, local optimality does not guarantee global optimality.
	Hence we run local optmizations from starting points on a grid with mesh side $0.2$ over the subspace $[0, 5]^4$ of the parameter space, and select our global maximum candidate as the local maximum with the highest entropy rate per unit time.
	Even though we cannot rigorously prove that our global maximum candidate $M_{\text{max}}=M(\vec{u},\tens{B}_{\text{max}})$ as represented by the red dot in Fig.~\ref{fig:model_id} with
	\begin{equation} 
		\tens{B}_{\text{max}} \approx 
		\begin{pmatrix} -2.723 & 1.821 \\ 1.098 & -2.277 \end{pmatrix}\,\yr^{-1}
	\end{equation}
	and $\theta_{\text{max}}\approx 1.916$ is a global maximum, we can clearly see that it is a good candidate.
	Increasing distance of local maximum parameters (panel (a)) and mean transit time (b) from the global maximum candidate lead to a decrease in entropy rate per unit time.
	Furthermore, local optimizations with starting points on the grid lead only to small improvements. A good choice of starting point on the grid is crucial to find a good global maximum candidate (c).
	Finally, the global maximum candidate for the entropy rate per unit time does not maximize the path entropy (d).
	
\section{Discussion}
Based on the stochastic path that a single particle takes through a deterministic compartmental system, we introduced three types of entropy based on Shannon's information theory.
The entropy of the particle's entire path through the system is the central concept, and the entropy rates per unit time and per jump are consistently derived from it.
Even though we call $\H(\mathcal{P})$ path entropy and identify models by maximizing it, it is different from the concept of path entropy as treated in the context of maximum caliber (MaxCal) \citep{jaynes1985macroscopic, Roach2020}.
We maximize here the Shannon entropy of a single particle's microscopic path through a compartmental system by means of an absorbing continuous-time Markov chain whose transition probabilities are already determined by the macroscopic equilibrium state of the system.
As discussed by \citet{Presse2013RMP}, MaxCal interprets the path entropy as a macroscopic system property to be maximized in order to identify a time-dependent trajectory of the entire dynamical system, not just one single particle.

In the field of soil carbon cycle modeling, \citet{Agren2021BGC} applied the maximum entropy principle to identify the distribution of soil carbon qualities within the framework of the continuous-quality theory.
Given only the nonnegative mean quality, an application of MaxEnt leads to an exponential quality distribution, because under these circumstances the exponential distribution is the maximum entropy distribution.
The path entropy generalizes this approach to several interconnected compartments and jumps between them, while each sojourn time in a compartment is exponentially distributed.

From the simple examples in Sect.~\ref{sec:simple_examples} we can see that models can be ordered differently in terms of uncertainty, depending on whether the interest is in the uncertainty of the entire path or in some average uncertainty rate.
For applications of MaxEnt without restrictions on the transit time, it is often useful to maximize an entropy rate because by slowing the system down more and more, the path entropy can potentially be increased indefinitely and there is no way to find a maximum path entropy model.

By virtue of its very mathematical definition (Eq.~\eqref{eqn:entropy}), entropy is maximized when the system's symmetry is maximized.
This is indicated by the Bernoulli entropy (Fig.~\ref{fig:simple_entropy}, panel (a)) and supported by Example~\ref{max_ent_example_1}.
Intuitively, this result is obvious.
If a system has high symmetry, a particle is equally likely to jump among different pools.
The Poisson process with intensity rate $1$ is the one with maximum entropy rate, which follows directly from properties of the function $f(x) = x,\log x$.
Furthermore, the resulting rates $z_j = 1/\E\left[\TT\right]$ of leaving the system are chosen such that the mean transit time constraint is fulfilled. 
In Example~\ref{max_ent_example_2}, the symmetry is broken by the additional restriction of a given steady-state vector.
Consequently, $\H(\mathcal{P}^\ast_2) \leq \H(\mathcal{P}^\ast_1)$.

When we compute entropy values for actual carbon-cycle models (Sects.~\ref{sec:example_1} and~\ref{sec:example_2}), we note that environmental or eco-physiological factors might impact model entropies.
For example, higher global surface temperatures might induce a higher global carbon cycle system speed ($1 < \xi << 6$).
This higher system speed reduces the uncertainty of the long-term future of entire paths of carbon atoms entering the terrestrial biosphere from the atmosphere.
At the same time, it increases the entropy rate per unit time, \ie, the uncertainty of the short-term future of carbon atoms already in the terrestrial biosphere.

Furthermore, we see that for sufficiently fast systems, a multi-pool model has lower entropy than a one-pool model with the same system speed.
The one-pool system might put too much weight on the uncertainties of a small amount of slow-cycling particles, while the more detailed multi-pool model focuses more on the small uncertainties of the major amount of fast-cycling particles.
The path of a detailed model that separates fast from slow paths is then even easier to predict than a one-pool model path, even though the detailed model's path looks more complicated.
However, detailed paths of slow-cycling systems are harder to predict than just the exit-time in a one-pool equivalent.

The two carbon-cycle models (Sects.~\ref{sec:example_1} and~\ref{sec:example_2}) are well understood in equilibrium, hence they can serve as a means to better understand properties of the newly introduced entropy metrics.
Once we understand entropy properties in dependence on general system properties, we can extrapolate this understanding to far more complex systems and make qualitative statements about their predictability without going into all model details.
One major insight from those two examples is that, in general, slow heterogeneous systems are much harder to predict than fast homogeneous systems.
Slowness increases the uncertainty of the duration of particle's stay in the system.
Heterogeneity increases the uncertainty of a particle's sequence of visited compartments.

These simple insights allow us to understand modeling issues on grander scale, like the huge difference in the diversity of modeling approaches for carbon uptake by photosynthesis and the carbon cycle in soils.
Both photosynthesis \citep{garcia2022mathematical} and soil carbon turnover \citep{manzoni2009soil} are modeled by many different approaches.
However, in ecosystem models photosynthesis is almost exclusively represented based \citep{zaehle2014evaluation} on the Farquhar model \citep{farquhar1980biochemical}, while soil carbon dynamics are represented by a great variety of models with very different structures \citep{friedlingstein2006climate}.
The latter leads to large variations in the prediction of future land carbon uptake \citep{friedlingstein2006climate,friedlingstein2014uncertainties}.
A comparison of carbon simulations from eleven model centers showed that across models global soil carbon varied more than twice as much as global net primary productivity \citep{todd2013causes}.
Leaves have evolved to serve a specific purpose: to take up carbon from the atmosphere.
Soils, on the other hand, have not been built to serve a specific purpose.
They have evolved as a dumpster for material of which every soil agent tries to take advantage.
This increases the heterogeneity of biogeochemical processes taking places in the soil.
Furthermore, soil carbon turnover happens on much larger time scales than photosynthesis.
While the photosynthetic apparatus operates on time scales in the order of split seconds to minutes, soil carbon cycling rates are in the order of decades to millenia.
Consequently, the higher uncertainty of soil carbon cycling compared to photosynthetic carbon uptake is an inherent property of the system.
Simply by the soil's heterogeneous and slow-cycling nature, the system has high inherent uncertainty, which hints at a theoretical limit that cannot be overcome by any model.

The example of model identification by MaxEnt in Sect.~\ref{sec:model_identification} shows a major difference to the more artificial previous maximum entropy examples.
The given constraints do not tell us enough about the structure of the model class $\mathcal{M}$ to ensure that an identified local maximum is also a global maximum.
Owing to the nonlinear restrictions on the parameters in Eq.~\eqref{eqn:measurement_data}, the parameter space is probably not convex.
Hence, local maxima are not guaranteed to be also globally optimal.
The small size system size allows us to identify a reasonable global maximum candidate model by brute force, starting local maximizations on a grid over a parameter sub-space.
Practical examples might include higher-dimensional systems and thus not be feasible for brute-force approaches.
More sophisticated optimizations methods suitable for the particular problem at hand should then be applied.

\section{Conclusions}
The information content of autonomous compartmental systems in equilibrium can be assessed by the entropy of the path of particles traveling through the system of interconnected compartments. When a particle moves through a compartmental system, it creates a path from the time of its entry until the time of its exit. This path can be described in three ways: (1) as a random variable in the path space; (2) as a continuous-time stochastic process representing the visited compartments; (3) as a discrete sequence of pairs consisting of visited compartments and associated sojourn times. 
Based on these three possible descriptions, we introduced for systems in equilibrium (1) the entropy of the entire path, (2) the entropy rate per unit time, and (3) the entropy rate per jump. These three different entropies allow us to quantify how difficult it is to predict the path of particles entering a compartmental system, serving as a measure of system uncertainty/predictability. With these measures it is thus possible to apply maximum entropy principles to compartmental systems in equilibrium in order to address problems of equifinality in model selection. 

Although the path entropy concept developed here only applies to systems in equilibrium, it sets the foundation for future research on systems out of equilibrium. 
This could be done by building on the concept of the entropy rate per unit time as an instantaneous uncertainty and interpreting non-autonomous compartmental systems as inhomogeneous Markov chains.
This would allow an extension of MaxCal so far applied only to the inhomogeneous embedded jump chain as done by \citet{Ge2012JCP} to incorporate also sojourn times in different compartments.

By introducing the concept of path entropy to compartmental systems, we made a first crucial step toward a quantification of information content in models that can be compared to other methods to obtain information content from observations. Using entropy measures in both models and observations, we could potentially advance toward better methods for model selection applying the maximum entropy principle. 

\newpage

\begin{acknowledgements}
Funding was provided by the Max Planck Society and the German Research Foundation through its Emmy Noether Program (SI 1953/2--1) and the Swedish Research Council for Sustainable Development FORMAS, under grant 2018-01820.
\end{acknowledgements}

\bibliographystyle{MG}
{footnotesize
\bibliography{entropy.bib}}

\appendix

\section{Proves of the MaxEnt examples}
	Recall that the path entropy of a linear autonomous compartmental system $M=M(\vec{u},\tens{B})$ is given by
	\begin{equation}
    \begin{aligned}
      \H(\mathcal{P}(M)) &= \H(X)\\
      &= -\suml_{i=1}^d\beta_i\,\log\beta_i + \suml_{j=1}^d \frac{x^\ast_j}{\vnorms{\vec{u}}}\left[\suml_{i=1,i\neq j}^d \,B_{ij}\,(1-\log B_{ij}) + z_j\,(1-\log z_j)\right].
    \end{aligned}
	\end{equation}
	In order to obtain maximum entropy models under simple constraints, we now adapt ideas of \cite{Girardin2004MCAP}.

	\begin{myproposition}
    \label{proposition:max_ent_example_1}
		Consider the set $\mathcal{M}_1$ of compartmental systems in equilibrium given by Eq.~\eqref{eqn:lin_CS_sys} with a predefined nonzero input vector $\vec{u}$, a predefined mean transit time $\E\left[\TT\right]$, and an unknown steady-state vector comprising nonzero components.
		The compartmental system $M^\ast_1=M(\vec{u},\tens{B}^\ast)$ with 
		\begin{equation}
			\tens{B}^\ast = \begin{pmatrix}
									-\lambda & 1 & \cdots & 1\\
									1 & -\lambda & 1 \cdots & 1 \\
									\vdots & & \ddots & \vdots\\
									1 & \cdots & 1 & -\lambda
             \end{pmatrix},
		\end{equation}
		where $\lambda=d-1+1/\E\left[\TT\right]$, 		
		is the maximum entropy model in $\mathcal{M}_1$.
	\end{myproposition}

	\begin{proof}
		We can express the constraint $\E\left[\TT\right] =\vnorms{\vec{x}^\ast}/\vnorms{\vec{u}}$ by
		\begin{equation}
		C_1 = \frac{1}{\vnorms{\vec{u}}}\,\suml_{j=1}^d x_j^\ast - \E\left[\TT\right] = 0.
		\end{equation}
		From the steady-state formula $\vec{x}^\ast = -\tens{B}^{-1}\,\vec{u}$, we obtain another set of $d$ constraints, which we can describe by
		\begin{equation}
			\frac{1}{\vnorms{\vec{u}}}\,(\tens{B}\,\vec{x}^\ast)_i=-\beta_i,\quad i=1,2,\ldots,d.
		\end{equation}
		We rewrite the left hand side as
		\begin{equation}
      \begin{aligned}
        \frac{1}{\vnorms{\vec{u}}}\,(\tens{B}\,\vec{x}^\ast)_i &= \frac{1}{\vnorms{\vec{u}}}\,\suml_{j=1}^d B_{ij}\,x_j^\ast = \frac{1}{\vnorms{\vec{u}}}\,\left(\suml_{j=1,j\neq i}^d B_{ij}\,x_j^\ast + B_{ii}\,x_i^\ast\right)\\
        &= \frac{1}{\vnorms{\vec{u}}}\,\suml_{j=1,j\neq i}^d B_{ij}\,x_j^\ast-\frac{1}{\vnorms{\vec{u}}}\,x_i^\ast\,\left(\suml_{k=1,k\neq i}^d B_{ki} + z_i\right), 
      \end{aligned}
    \end{equation}
		which leads to the constraints
		\begin{equation}\label{eqn:constraint_C2}
			C_{2,i} = \frac{1}{\vnorms{\vec{u}}}\,\suml_{j=1,j\neq i}^d B_{ij}\,x_j^\ast-\frac{1}{\vnorms{\vec{u}}}\,x_i^\ast\,\left(\suml_{k=1,k\neq i}^d B_{ki} + z_i\right) + \beta_i = 0,\quad i\in S.
		\end{equation}	
		The Lagrangian is now given by
		\begin{equation}\label{eqn:Lagrangian}
			L = \H(X) + \gamma_0\,C_1 + \suml_{i=1}^d \gamma_i\,C_{2,i}
		\end{equation}
		and its partial derivatives with respect to $B_{ij}\,(i\neq j)$, $z_j$, and $ x_j^\ast$ by
    \begin{equation}
      \begin{aligned}
        \vnorms{\vec{u}}\,\pderiv{B_{ij}}\,L &=  -x_j^\ast\,\log B_{ij} +   \gamma_i\,x_j^\ast- \gamma_j\,x_j^\ast,\\
        \vnorms{\vec{u}}\,\pderiv{z_j}\,L &= -x_j^\ast\,\log z_j-\gamma_j\,x_j^\ast,
      \end{aligned}
    \end{equation}
		and
		\begin{equation}
      \begin{aligned}
        \vnorms{\vec{u}}\,\pderiv{x_j^\ast}\,L &= \suml_{i=1,i\neq j}^d B_{ij}\,(1-\log B_{ij}) + z_j\,(1-\log z_j)\\
        &\quad+ \gamma_0 + \suml_{i=1,i\neq j}^d \gamma_i\,B_{ij} - \gamma_j\,\left(\suml_{k=1,k\neq j}^d B_{kj} + z_j\right), 
      \end{aligned}
    \end{equation}
		respectively.
		Setting $\pderiv{B_{ij}}\,L=0$ gives $B_{ij} = e^{\gamma_i-\gamma_j}$, and setting $\pderiv{z_j}\,L = 0$ gives $z_j = e^{-\gamma_j}$.
		We plug this into $\pderiv{x_j^\ast}\,L=0$ and get
		\begin{equation}
      \begin{aligned}
        0 &= \suml_{i=1,i\neq j}^d e^{\gamma_i-\gamma_j}\,[1-(\gamma_i-\gamma_j)] + e^{-\gamma_j}\,[1-(-\gamma_j)]\\
        &\quad + \gamma_0 + \suml_{i=1,i\neq j}^d \gamma_i\,e^{\gamma_i-\gamma_j} - \gamma_j\,\left(\suml_{k=1,k\neq j}^d e^{\gamma_k-\gamma_j}+e^{-\gamma_j}\right)\\
        &= \suml_{i\neq j,i\neq j} e^{\gamma_i-\gamma_j}+e^{-\gamma_j} + \gamma_0.
      \end{aligned}
    \end{equation}
		Subtracting $e^{-\gamma_j}$ from both sides and multiplying with $e^{\gamma_j}$ leads to
		\begin{equation}
			\gamma_0\,e^{\gamma_j}+\suml_{i=1,i\neq j}^d e^{\gamma_i} = -1,\quad j=1,2,\ldots,d.
		\end{equation}
		This is equivalent to the linear system $\tens{Y}\,\vec{v} = \vec{-1}$ with
		\begin{equation}
			\tens{Y} = \begin{pmatrix}
									\gamma_0 & 1 & \cdots & 1\\
									1 & \gamma_0 & 1 \cdots & 1 \\
									\vdots & & \ddots & \vdots\\
									1 & \cdots & 1 & \gamma_0
								\end{pmatrix},\quad
			\vec{v} = \begin{pmatrix} e^{\gamma_1}\\ e^{\gamma_2}\\ \vdots \\ e^{\gamma_d} \end{pmatrix},\quad
			\vec{-1} = \begin{pmatrix} -1 \\ -1 \\ \vdots \\ -1 \end{pmatrix}.
		\end{equation}
		The case $\gamma_0=1$ has no solution $\vec{v}$ since $e^{\gamma_i}>0>-1$.
		For $\gamma_0\neq 1$ the matrix $\tens{Y}$ has a nonzero determinant which makes the system uniquely solvable.
		For symmetry reasons, $\gamma_i=\gamma_j=:\gamma$ for all $i,j=1,2,\ldots,d$.
		Consequently, for $i\neq j$, we get $B_{ij}=1$, and by summing Eq.~\eqref{eqn:constraint_C2} over $i\in S$,
		\begin{equation}
		\begin{aligned}
			0 &= \vnorms{\vec{u}}\,\suml_{i=1}^d C_{2,i} = \suml_{i=1}^d \suml_{j=1,j\neq i}^d B_{ij}\,x_j^\ast - \suml_{i=1}^d x_i^\ast\,\left(\suml_{k=1,k\neq i}^d B_{ki}+z_i\right) - \vnorms{\vec{u}}\\
			&= -\suml_{i=1}^d x_i^\ast\,z_i - \vnorms{\vec{u}},		
		\end{aligned}
		\end{equation}
		which can also be expressed by $\vec{z}^T\,\vec{x}^\ast = \vnorms{\vec{u}}$.
		We simply plug in $z_i=e^{-\gamma}$ and get $e^{-\gamma}\,\vnorms{\vec{x}^\ast} = \vnorms{\vec{u}}$, which means $z_i = 1/\E\left[\TT\right]$.
		Consequently,
		\begin{equation}
			\tens{B}^\ast = \begin{pmatrix}
									-\lambda & 1 & \cdots & 1\\
									1 & -\lambda & 1 \cdots & 1 \\
									\vdots & & \ddots & \vdots\\
									1 & \cdots & 1 & -\lambda
							\end{pmatrix}
		\end{equation}
		for $\lambda=d-1+1/\E\left[\TT\right]$.
		Since uniqueness of this solution follows from its construction, we remain with showing maximality.
		To this end, we split the entropy into to three parts, \ie, $\H(X) = H_1 + H_2 + H_3$ with
		\begin{equation}
			\begin{aligned}
				H_1 &= -\suml_{i=1}^d\beta_i\,\log\beta_i,\\
				H_2 &= \suml_{j=1}^d \frac{x^\ast_j}{\vnorms{\vec{u}}}\,z_j\,(1-\log z_j), \text{ and}\\
				H_3 &= \suml_{j=1}^d \frac{x^\ast_j}{\vnorms{\vec{u}}}\,\suml_{i=1,i\neq j}^d \,B_{ij}\,(1-\log B_{ij}).
			\end{aligned}
		\end{equation}
		The term $H_1$ is independent of $B_{ij}$ and $z_j$ for all $i,j\in S$ and $i\neq j$, and can thus be ignored.
		We denote by $E$ the pool from which the particle exits from the system.
		Then we can use \citep[Sect.~5.3]{Metzler2018MGS}
		\begin{equation}
		  \P(E=j) = \frac{z_j\,x^\ast_j}{\vnorms{\vec{u}}}
		\end{equation}
    to rewrite the second term as
		\begin{equation}
			H_2 = \suml_{j=1}^d \P(E=j)\,(1-\log z_j) = \suml_{j=1}^d \H(T_E\,|\,E=j)\,\P(E=j) = \H(T_E\,|\,E),
		\end{equation}
		where $T_E$ denotes the exponentially distributed sojourn time in $E$ right before absorption.
		We see that $H_2$ becomes maximal if the knowledge of $E$ contains no information about $T_E$.
		Hence, $z_j=z_i$ for $i,j\in S$.
		Since we need to ensure the systems' constraint on $\E\left[\TT\right]$, we get $z_j=1/\E\left[\TT\right]$ for all $j\in S$.
		
		In order to see that $B_{ij}=1$ ($i\neq j$) leads to maximal entropy, we first note that 
		\begin{equation}
			H_3 = \suml_{j=1}^d \frac{x^\ast_j}{\vnorms{\vec{u}}}\,\suml_{i=1,i\neq j}^d \,1\cdot(1-\log 1) = (d-1)\,\suml_{j=1}^d \E\left[O_j\right] = (d-1)\,\E\left[\TT\right]
		\end{equation}
		by Eq.~\eqref{eqn:H_occupation_time}.
		We now disturb $B_{kl}$ for fixed $k,l\in S$ with $k\neq l$ by a sufficiently tiny $\varepsilon$, which may be positive or negative.
		We define $B_{kl}(\varepsilon):=B_{kl}+\varepsilon$, and a change from $\lambda_j$ to $\lambda_j(\varepsilon):=\lambda_j+\varepsilon>0$ ensures $z_j(\varepsilon) = z_j$, implying that the system's mean transit time remains unchanged, \ie, $\E\left[\TT(\varepsilon)\right] = \E\left[\TT\right]$.
		The $\varepsilon$-disturbed $H_3$ is given by
		\begin{equation}
      \begin{aligned}
        H_3(\varepsilon) &= \suml_{j=1}^d \frac{x^\ast_j(\varepsilon)}{\vnorms{\vec{u}}}\,\suml_{i=1,i\neq j}^d \,1\cdot(1-\log 1)\,\left(1-\mathbbm{1}_{\{i=k,\,j=l\}}\right)\\
        &\quad + \frac{x^\ast_l(\varepsilon)}{\vnorms{\vec{u}}}\,(1+\varepsilon)\,[1-\log(1+\varepsilon)]\\
        &= \suml_{j=1}^d \frac{x^\ast_j(\varepsilon)}{\vnorms{\vec{u}}}\,\suml_{i=1,i\neq j}^d \,\left(1-\mathbbm{1}_{\{i=k,\,j=l\}}\right) + \frac{x^\ast_l(\varepsilon)}{\vnorms{\vec{u}}}\,(1-\delta)\\
      \end{aligned}
    \end{equation}
		for some $\delta>0$ since the map $x\mapsto x\,(1-\log x)$ has its global maximum at $x=1$.
		Consequently,
		\begin{equation}
		\begin{aligned}
			&H_3(\varepsilon) &&= \left[\suml_{j=1}^d \frac{x^\ast_j(\varepsilon)}{\vnorms{\vec{u}}}\,\suml_{i=1,i\neq j}^d \,1\right] - \delta\,\frac{x^\ast_l(\varepsilon)}{\vnorms{\vec{u}}}
			&&= (d-1)\,\suml_{j=1}^d \E\left[O_j(\varepsilon)\right] - \delta\,\frac{x^\ast_l(\varepsilon)}{\vnorms{\vec{u}}}\\
			& &&= (d-1)\,\E\left[\TT(\varepsilon)\right] - \delta\,\frac{x^\ast_l(\varepsilon)}{\vnorms{\vec{u}}}
			&&= (d-1)\,\E\left[\TT\right] - \delta\,\frac{x^\ast_l(\varepsilon)}{\vnorms{\vec{u}}}\\
			& && < H_3.
		\end{aligned}
	\end{equation}
		Hence, disturbing $B_{ij}$ away from $1$ reduces the entropy of the system, and the proof is complete.
	\end{proof}
	
	\begin{myproposition}
    \label{proposition:max_ent_example_2}
		Consider the set $\mathcal{M}_2$ of compartmental systems in equilibrium given by Eq.~\eqref{eqn:lin_CS_sys} with a predefined nonzero input vector $\vec{u}$ and a predefined positive steady-state vector $\vec{x}^\ast$.
		The compartmental system $M^\ast_2=M(\vec{u},\tens{B}^\ast)$ with $\tens{B}^\ast=(B_{ij})_{i,j\in S}$ given by
		\begin{equation}
			B_{ij} = \begin{cases}
							\sqrt\frac{x_i^\ast}{x_j^\ast},\quad & i\neq j,\\
							-\suml_{k=1,k\neq j}^d \sqrt\frac{x_k^\ast}{x_j^\ast} - \frac{1}{\sqrt{x_j^\ast}}, \quad &i=j,
						\end{cases}
		\end{equation}
		is the maximum entropy model in $\mathcal{M}_2$.
	\end{myproposition}

	\begin{proof}
		The mean transit time $\E\left[\TT\right]=\vnorms{\vec{x}^\ast}/\vnorms{\vec{u}}$ of the system is fixed.
		Hence, the Lagrangian $L$ is the same as in Eq.~\eqref{eqn:Lagrangian}, and setting $\partial L / \partial B_{ij} = 0$ leads  to
		\begin{equation}
			-\log B_{ij} + \gamma_i-\gamma_j = 0,\quad i\neq j.
		\end{equation}
		An interchange of the indices and summing the two equations gives
		\begin{equation}
			\log B_{ij} + \log B_{ji} = 0.
		\end{equation}
		Hence, $B_{ij}\,B_{ji} = 1$.
		A good guess gives $B_{ij}^2 = x_i^\ast/x_j^\ast$ and $\gamma_j = \frac{1}{2}\,\log x_j^\ast$.
		From $\pderiv{z_j}\,L=0$ we get
		\begin{equation}
			-\log z_j -\gamma_j = 0,\quad j\in S,
		\end{equation}
		and in turn $z_j=(x_j^\ast)^{-1/2}$.
		Maximality and uniqueness of this solution follow from the strict concavity of $\H(X)$ as a function of $B_{ij}$ and $z_j$ for fixed $\vec{x}^\ast$.
		We can see this strict concavity by 
		\begin{equation}
			\frac{\partial^2}{\partial B_{ij}^2}\,\H(X) = \pderiv{B_{ij}}\,\frac{-x_j^\ast}{\vnorms{\vec{u}}}\,\log B_{ij} = -\frac{x_j^\ast}{\vnorms{\vec{u}}\,B_{ij}} < 0
		\end{equation}
		and
		\begin{equation}
			\frac{\partial^2}{\partial z_j^2}\,\H(X) = \pderiv{z_j}\,\frac{-x_j^\ast}{\vnorms{\vec{u}}}\,\log z_j = -\frac{x_j^\ast}{\vnorms{\vec{u}}\,z_i} < 0.
		\end{equation}
	\end{proof}

\end{document}

%% file: figs/Compartments/onepool.tex
%

\begin{tikzpicture}
\node at (-1, 0) [shape=circle] (in) {};
\node at (0, 0) [shape=circle, draw] (x1) {$x_1$};
\node at (1, 0) [shape=circle] (out) {};
\draw [->] (in) to  (x1);
\draw [->] (x1) to (out);
\end{tikzpicture}


%% file: figs/Compartments/twoseries.tex
%

\begin{tikzpicture}
\node at (-1, 0) [shape=circle] (in) {};
\node at (0, 0) [shape=circle, draw] (x1) {$x_1$};
\node at (1, 0) [shape=circle, draw] (x2) {$x_2$};
\node at (2, 0) [shape=circle] (out) {};
\draw [->] (in) to  (x1);
\draw [->] (x1) to (x2);
\draw [->] (x2) to (out);
\end{tikzpicture}


%% file: figs/Compartments/twoparallel.tex
%

\begin{tikzpicture}
\node at (-1, 0.5) [shape=circle] (in1) {};
\node at (-1, -0.5) [shape=circle] (out1) {};
\node at (0, 0) [shape=circle, draw] (x1) {$x_1$};
\node at (1, 0) [shape=circle, draw] (x2) {$x_2$};
\node at (2, 0.5) [shape=circle] (in2) {};
\node at (2, -0.5) [shape=circle] (out2) {};
\draw [->] (in1) to  (x1);
\draw [->] (x1) to (out1);
\draw [->] (in2) to (x2);
\draw [->] (x2) to (out2);
\end{tikzpicture}


%% file: figs/Compartments/twofeedback1.tex
%

\begin{tikzpicture}
\node at (-1, 0.5) [shape=circle] (in1) {};
\node at (-1, -0.5) [shape=circle] (out1) {};
\node at (0, 0) [shape=circle, draw] (x1) {$x_1$};
\node at (1, 0) [shape=circle, draw] (x2) {$x_2$};
\node at (2, 0) [shape=circle] (out2) {};
\draw [->] (in1) to  (x1);
\draw [->] (x1) to (out1);
\draw [<->] (x1) to (x2);
\draw [->] (x2) to (out2);
\end{tikzpicture}


%% file: figs/Compartments/twofeedback2.tex
%

\begin{tikzpicture}
\node at (-1, 0.5) [shape=circle] (in1) {};
\node at (-1, -0.5) [shape=circle] (out1) {};
\node at (0, 0) [shape=circle, draw] (x1) {$x_1$};
\node at (1, 0) [shape=circle, draw] (x2) {$x_2$};
\node at (2, 0.5) [shape=circle] (in2) {};
\node at (2, -0.5) [shape=circle] (out2) {};
\draw [->] (in1) to  (x1);
\draw [->] (x1) to (out1);
\draw [<->] (x1) to (x2);
\draw [->] (in2) to (x2);
\draw [->] (x2) to (out2);
\end{tikzpicture}


%% file: figs/Compartments/threeseries.tex
%

\begin{tikzpicture}
\node at (-1, 0) [shape=circle] (in) {};
\node at (0, 0) [shape=circle, draw] (x1) {$x_1$};
\node at (1, 0) [shape=circle, draw] (x2) {$x_2$};
\node at (2, 0) [shape=circle, draw] (x3) {$x_3$};
\node at (3, 0) [shape=circle] (out) {};
\draw [->] (in) to  (x1);
\draw [->] (x1) to (x2);
\draw [->] (x2) to (x3);
\draw [->] (x3) to (out);
\end{tikzpicture}


%% file: figs/Compartments/threeparallel.tex
%

\begin{tikzpicture}
\node at (-1, 0.5) [shape=circle] (in1) {};
\node at (-1, -0.5) [shape=circle] (out1) {};
\node at (0, 0) [shape=circle, draw] (x1) {$x_1$};
\node at (1, 0) [shape=circle, draw] (x2) {$x_2$};
\node at (2, 0) [shape=circle, draw] (x3) {$x_3$};
\node at (0, 0.5) [shape=circle] (in2) {};
\node at (2, -0.5) [shape=circle] (out2) {};
\node at (3, 0.5) [shape=circle] (in3) {};
\node at (3, -0.5) [shape=circle] (out3) {};
\draw [->] (in1) to  (x1);
\draw [->] (x1) to (out1);
\draw [->] (in2) to (x2);
\draw [->] (x2) to (out2);
\draw [->] (in3) to (x3);
\draw [->] (x3) to (out3);
\end{tikzpicture}
